\pdfoutput=1

\documentclass[%
 reprint,
 superscriptaddress,
 nobibnotes,
 amsmath,amssymb,
 aps,
 showkeys,
 longbibliography,
]{revtex4-2}

\usepackage{graphicx}% Include figure files
\usepackage{dcolumn}% Align table columns on decimal point
\usepackage{bm}% Bold math
\usepackage{chemformula}% Chemical formulas
\usepackage[separate-uncertainty=true,separate-uncertainty,multi-part-units=single]{siunitx}% SI units with separate uncertainty
\usepackage{hyperref}% Add hypertext capabilities
\hypersetup{colorlinks=true,allcolors=magenta}

\begin{document}

\title{Reparameterizing GFN1-xTB for Ionic Liquids}

\author{F. Stavarache}
    \affiliation{Materials Simulation \& Modelling, Department of Applied Physics and Science Education, Eindhoven University of Technology, 5600 MB, Eindhoven, The Netherlands}
\author{S. Calero}
    \affiliation{Materials Simulation \& Modelling, Department of Applied Physics and Science Education, Eindhoven University of Technology, 5600 MB, Eindhoven, The Netherlands}
    \affiliation{Eindhoven Institute for Renewable Energy Systems (EIRES), Eindhoven University of Technology, Eindhoven 5600 MB, The Netherlands}
\author{J. M. Vicent-Luna}
    \email[Corresponding author: ]{j.vicent.luna@tue.nl}
    \affiliation{Materials Simulation \& Modelling, Department of Applied Physics and Science Education, Eindhoven University of Technology, 5600 MB, Eindhoven, The Netherlands}
    \affiliation{Eindhoven Institute for Renewable Energy Systems (EIRES), Eindhoven University of Technology, Eindhoven 5600 MB, The Netherlands}

\date{\today}

\begin{abstract}

Ionic liquids are highly tunable materials whose properties can be tailored for a wide range of applications, but their chemical diversity and complex electronic structure make systematic screening using quantum-mechanical methods computationally demanding. Semiempirical quantum methods offer an efficient alternative, although their accuracy depends strongly on the underlying parameterization. In this work, we investigate the reparameterization of the density functional tight-binding method GFN1-xTB for predicting structural, energetic, and electronic properties of ionic liquids. For this purpose, we built a database containing over 14,000 conformations optimized using classical, semiempirical, and density functional theory calculations. We used this database to construct training datasets and explored different parameterization strategies fitted to binding energies, forces, HOMO-LUMO gaps, and geometries. Analysis of the original GFN1-xTB model reveals systematic errors in these properties, particularly in HOMO-LUMO gaps, with distinct behavior for saturated and aromatic cations. Based on this analysis, we investigate two models: CXMIM-xTB, a targeted model trained on imidazolium-based ionic liquids, and IL-xTB, a more general model designed to describe systems containing both saturated and aromatic cations. CXMIM-xTB substantially improves the description of binding energies, geometries, and frontier orbital energies while retaining the computational efficiency of GFN1-xTB. The model is subsequently applied to estimate electrochemical windows of pure and mixed ionic liquids, providing values consistent with DFT-based approaches and experimental ranges. IL-xTB follows a hierarchical parameterization strategy including different cation and anion families and substantially reduces the systematic errors of GFN1-xTB across the chemical space considered, providing a route for developing xTB parameterizations with broader applicability to ionic liquids. Overall, both parameterization strategies outperform the original GFN1-xTB model for specific tasks, demonstrating that reparameterized xTB methods provide an efficient approach for connecting configurational sampling with quantum-mechanical predictions of structural, energetic, and electronic properties in complex ionic systems.

\end{abstract}


\maketitle

%%%MAIN TEXT%%%%%%%%%%%%%%%%%%%%%%%%%%%%%%%%%%%%%%%%%%%%
%%%%%%%%%%%%%%%%%%%%%%%%%%%%%%%%%%%%%%%%%%%%%%%%%%%%%%%%
%%%%%%%%%%%%%%%%%%%%%%%%%%%%%%%%%%%%%%%%%%%%%%%%%%%%%%%%
%%%%%%%%%%%%%%%%%%%%%%%%%%%%%%%%%%%%%%%%%%%%%%%%%%%%%%%%

\section{Introduction}
\label{sec:intro}
%%%

Room-temperature Ionic Liquids (or simply Ionic Liquids, ILs), sometimes also called ionic melts, liquid salts, or ionic fluids, are soft materials made up of ions and can serve a wide range of applications. Because of their low vapor pressures, high thermal and electrochemical stability, and low melting points, they can be used as solvents for organic and catalytic reactions \cite{NANDA20211234, YANG200519, Thomas2016-ux, doi:10.1021/acssuschemeng.6b00553}. In addition, due to their ion-transfer ability, ILs can also be used as electrolytes for batteries \cite{ZHENG2022100075, doi:10.1021/acs.cgd.2c00706, LIU2021320}. These distinct characteristics arise from the strong Coulombic interactions and the structural diversity of the constituent ions, which offer a high degree of designability to tailor ILs for specific applications. For this reason, they are sometimes referred to as "designer solvents" \cite{https://doi.org/10.1002/zaac.201700001}.

Unlike traditional salts, such as \ch{NaCl}, where the ions are small and attractive Coulomb interactions force the material into a solid phase under ambient conditions, ILs consist of large asymmetric ions that cannot be properly constrained by neighboring molecules, forming a viscous liquid. Moreover, unlike traditional solvents, ILs have very low saturation pressures, meaning that they remain in the liquid phase even under ultra-high vacuum (UHV) \cite{Bier20012010}. For reference, while H\textsubscript{2}O has a vapor pressure of 3 kPa at 297 K, \ch{[BuMeim][PF6]} has a vapor pressure of 100 pPa at the same temperature.

In addition to their role as solvents, ionic liquids have recently found use in advanced functional materials, including photonic and optoelectronic systems \cite{MAHATO2023101808}. For example, IL-based systems have been proposed as white-light emitters and fluorescent sensors \cite{C4CC08825E, C9TC05377H, doi:10.1021/acs.iecr.5b03947, doi:10.1021/jp109121h}. Furthermore, ILs are of particular interest in electrochemistry, where their composition makes them promising electrolytes for lithium-ion batteries \cite{KOECH2024321}. Compared to commonly used salts such as \ch{Li[PF6]}, \ch{Li[BF4]}, or \ch{Li[ClO4]}, ILs are generally thermodynamically stable, reducing the risk of fire or explosions \cite{RANA2023116340}. However, their high viscosity can hinder ion transport, requiring careful tuning of their composition to optimize performance \cite{https://doi.org/10.1002/anie.202305331, KOECH2024321}.

An important advantage of ILs is the ability to tune their properties by combining different cations and anions in various ratios. While some properties, such as density and viscosity, follow predictable mixing rules, others exhibit nontrivial behavior \cite{C2CS35177C}. For example, combining \ch{[EtMeim][fsi]} with \ch{[EtMeim][BF4]} and \ch{[EtMeim][tfsi]} leads to a strong dependence of \ch{Li} intercalation reversibility on the \ch{[fsi]-} concentration \cite{SUGIMOTO2009802}. As a result, IL mixtures often require individual characterization, which can be both time-consuming and experimentally demanding.

A widely used alternative to experiments is the application of quantum chemical methods, particularly Density Functional Theory (DFT), to screen ionic mixtures and predict their properties. Both semi-local and hybrid functionals have been shown to accurately describe properties such as binding energies and HOMO-LUMO gaps in IL systems \cite{doi:10.1021/jp8107649, doi:10.1021/jp5034257}. However, the computational cost of DFT rapidly increases with system size. Since ionic pairs can easily contain more than 50 atoms and meaningful predictions require sampling multiple conformations, the application of DFT to realistic IL systems becomes computationally prohibitive.

Semiempirical quantum methods provide an attractive alternative by significantly reducing computational cost while retaining an electronic-structure description. Among these, Density Functional Tight Binding (DFTB) and its extended variants have gained considerable attention \cite{WOS:A1986C657800012, https://doi.org/10.1002/wcms.1493}. In particular, the GFN1-xTB method has become widely used due to its efficiency, robustness, and relatively simple parameterization \cite{doi:10.1021/acs.jctc.7b00118}. An important feature of such methods is that their accuracy can be systematically improved through reparameterization. Previous studies have demonstrated that targeted parameterization can significantly improve performance for specific classes of systems, including organosilicon compounds \cite{doi:10.1021/acs.jcim.1c01170}, atmospheric molecular clusters \cite{D4RA03021D}, and halide perovskites \cite{doi:10.1021/acs.jpcc.2c02412}. 

Despite these advances, a dedicated parameterization of GFN1-xTB for ionic liquids is still lacking. This is particularly relevant because ILs present a challenging environment for approximate quantum methods, where strong electrostatic interactions, charge delocalization, and structural diversity can lead to significant errors in electronic properties.

In this work, we investigate parameterization strategies to improve the performance of GFN1-xTB for ionic liquids. The focus is placed on improving the description of molecular orbital energies and optimized geometries, which are key quantities for the characterization of complex ionic systems. To this end, a comprehensive dataset of ionic conformations is constructed and used to train modified parameter sets. The resulting models are evaluated against DFT reference calculations, and their performance is assessed for both structural and electronic properties. This approach provides insight into the capabilities and limitations of targeted parameterization strategies for extending semiempirical quantum methods to ionic molecular systems.

\section{Methodology}
\label{sec:methods}
%%%

\begin{figure*}[t!]
    \centering
    \includegraphics[width=1.0\linewidth]{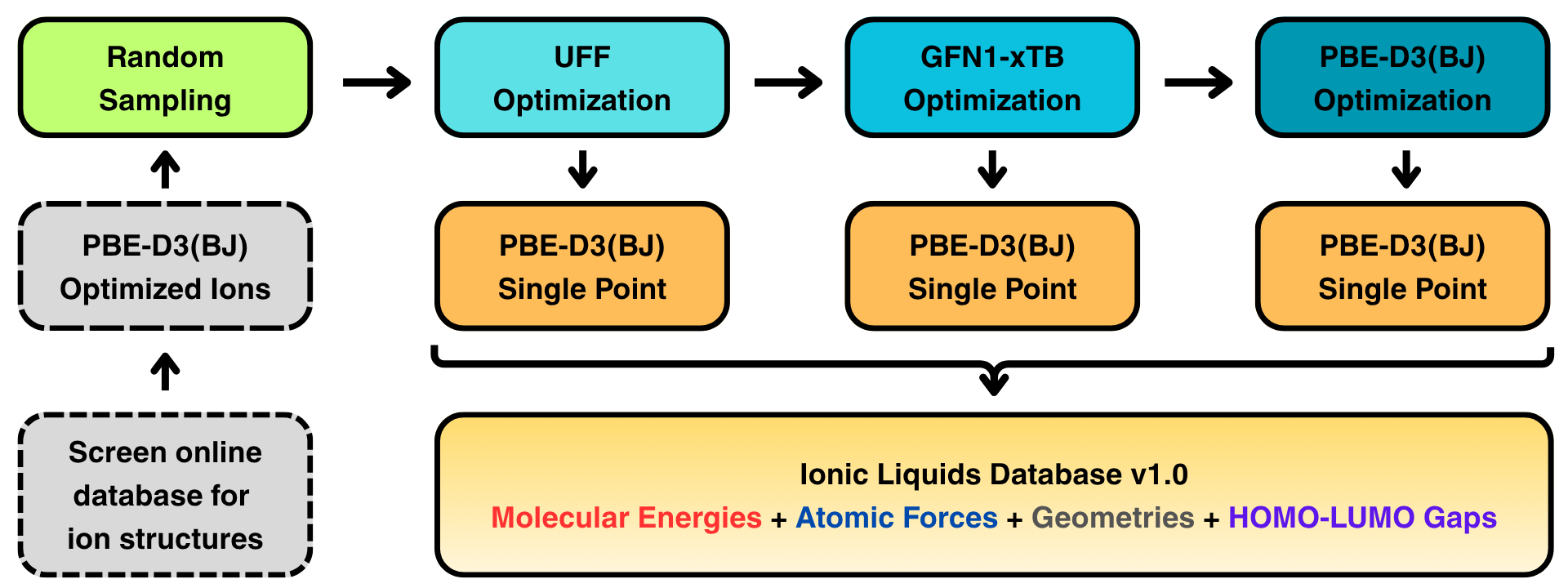}
    \caption{Workflow for generating the ionic-liquid dataset. Ion structures were sourced from PubChem and individually optimized at the PBE-D3(BJ) level. Randomly generated ion pairs were then processed through successive UFF, GFN1-xTB, and PBE-D3(BJ) geometry optimizations. Single-point PBE-D3(BJ) calculations were finally used to extract the reference properties employed during parameterization.}
    \label{fig:Fig_01}
\end{figure*}

A dedicated dataset was generated for this work because no publicly available dataset currently provides ionic-liquid structures and reference properties suitable for GFN1-xTB parameterization. Existing machine learning studies on ionic liquids have focused mainly on predicting thermodynamic properties from chemical composition rather than on producing structural datasets for atomistic model development \cite{LIU2022110888, LAAKSO2025121226}. Here, the term \textit{dataset} refers to the combination of molecular geometries and the reference properties extracted from them. The overall workflow used for data generation is shown in Figure \ref{fig:Fig_01}.

\begin{figure*}[t!]
    \centering
    \includegraphics[width=1\linewidth]{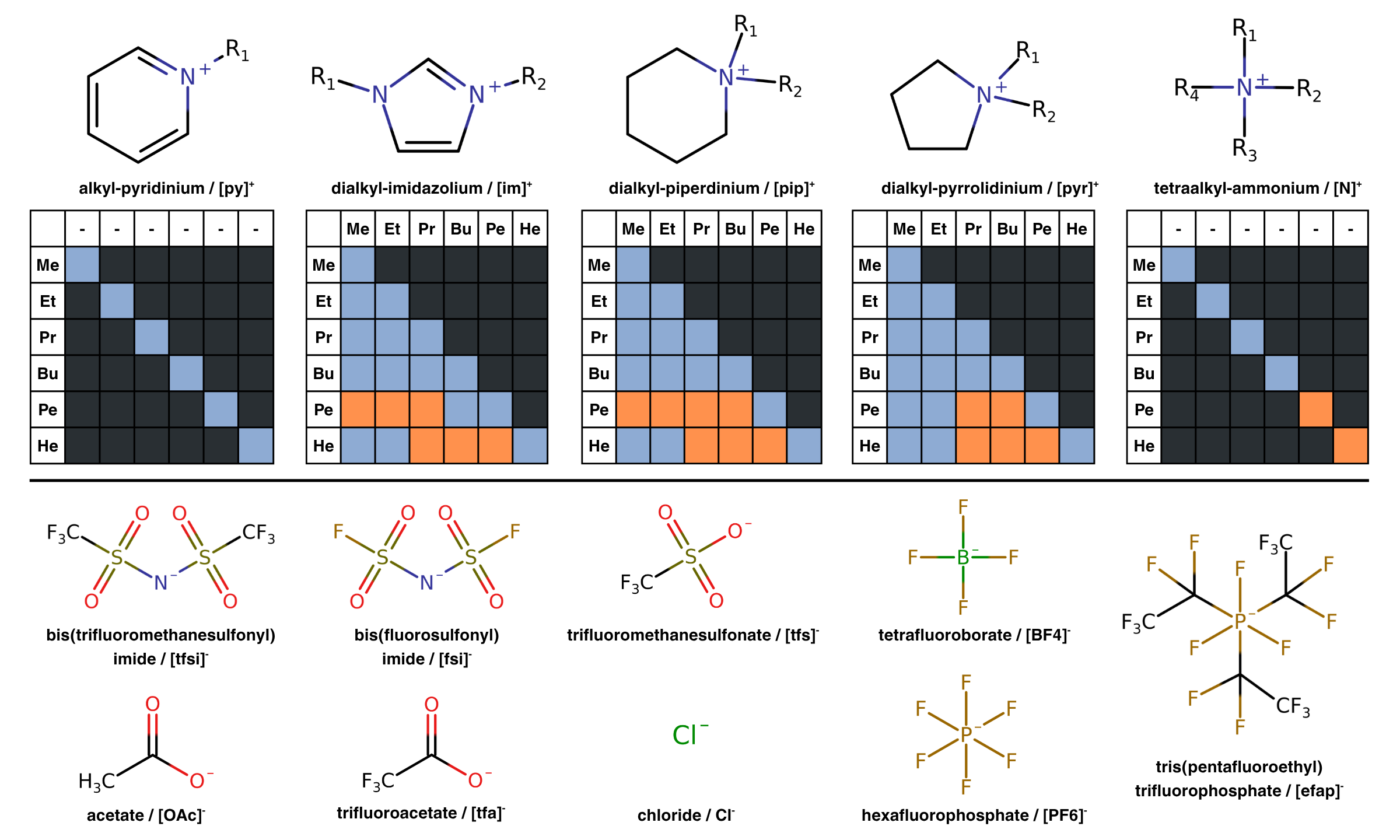}
    \caption{Chemical space considered in this work. The figure summarizes the cation families and anions included in the dataset. The square matrix shows the alkyl-chain combinations considered for each family, and the set of anions used to construct ionic pairs. Blue boxes represent alkyl-chains included in the dataset, while the orange boxes represent missing entries. For tetraalkyl-ammonium all chains have the same length. Ring-containing cations include dialkyl-imidazolium, dialkyl-piperidinium, dialkyl-pyrrolidinium, and alkyl-pyridinium species, while tetraalkyl-ammonium cations were included as saturated non-cyclic systems.}
    \label{fig:Fig_02}
\end{figure*}

The first step consisted of selecting ionic species that span a broad range of shapes and sizes without introducing excessive chemical diversity. This balance was achieved by focusing on nitrogen-containing cations built from a limited number of recurring structural motifs, combined with systematic variation of the alkyl side chains. The cation families and anions included in the dataset are summarized in Figure \ref{fig:Fig_02}.

To maintain reduced chemical complexity, only cations containing \ch{C}, \ch{H}, and \ch{N} atoms were selected. These include the saturated dialkyl-piperidinium (\textit{pip}) and dialkyl-pyrrolidinium (\textit{pyr}), the aromatic dialkyl-imidazolium (\textit{im}) and alkyl-pyridinium (\textit{py}), and tetraalkyl-ammonium (\ch{[N_{a,b,c,d}]+}) cations. This selection covers both aromatic and saturated motifs commonly found in ionic-liquid research while limiting the number of atomic species that need to be parameterized. Although additional cation classes could be considered in future work, the present choice captures a chemically meaningful subset of nitrogen-based ionic liquids.

The length of the alkyl chains was varied to include different balances of electrostatic and van der Waals interactions, which are known to affect the hydrophobic behavior of ionic liquids \cite{doi:10.1021/ja067427b, C3CP53035C}. For the ring-containing cations, alkyl substituents up to six carbon atoms were considered, although some combinations were excluded when no suitable PubChem structure was available or when the corresponding systems were omitted to limit the computational cost. For tetraalkyl-ammonium, only equal chain lengths up to four carbon atoms were included. In total, 55 cations were considered.

The cation structures were obtained from the PubChem database \cite{10.1093/nar/gkae1059}. When multiple conformations were available, one conformation per ion was selected in order to maximize the number of distinct ionic species included in the dataset. This choice limits the conformational variety of the isolated ions, but this limitation is partially compensated by the subsequent generation of multiple ion-pair arrangements.

The anion set was selected to provide chemically relevant diversity while keeping the number of species manageable. Nine anions were considered: chloride (\ch{Cl-}), tetrafluoroborate (\ch{[BF4]-}), hexafluorophosphate (\ch{[PF6]-}), acetate (\ch{[OAc]-}), trifluoroacetate (\ch{[tfa]-}), trifluoromethanesulfonate (\ch{[tfs]-}), bis(fluorosulfonyl)imide (\ch{[fsi]-}), bis(trifluoromethanesulfonyl)imide (\ch{[tfsi]-}), and tris(pentafluoroethyl)trifluorophosphate (\ch{[efap]-}). This set introduces \ch{P}, \ch{B}, \ch{Cl}, \ch{N}, \ch{C}, \ch{O}, \ch{S}, \ch{F}, and \ch{H} atoms into the dataset. Similar to the cations, one conformation per anion was used.

The isolated ions were first optimized at the PBE level of theory \cite{PhysRevB.54.16533} with D3 dispersion corrections \cite{10.1063/1.3382344} and Becke-Johnson damping \cite{https://doi.org/10.1002/jcc.21759}, hereafter denoted PBE-D3(BJ). These calculations were performed with the ADF engine \cite{https://doi.org/10.1002/jcc.1056} in AMS 2024.1 \cite{10.1063/5.0258496} using DZP basis sets \cite{https://doi.org/10.1002/jcc.10255}. Geometry optimizations converged when energy changes become smaller than $10^{-6}$ Ha, maximum gradient elements smaller than $10^{-4}$ Ha/\AA, RMS of gradients smaller than $2/3 \ 10^{-4}$ Ha/\AA, maximum step elements smaller than $10^{-3}$ \AA, and RMS of steps smaller than $2/3 \ 10^{-3}$ \AA. Although plain PBE is known to be limited for ionic liquids \cite{doi:10.1021/jp8107649}, the inclusion of D3(BJ) corrections provides a practical reference level for the structural and electronic properties targeted in this work \cite{doi:10.1021/jp5034257}.

The optimized ions were then paired to generate ionic configurations representative of ionic-liquid local structure. For this purpose, an in-house code was used to generate ten random ion-pair conformations without the need to define a simulation box. Starting from a small displacement vector of 0.1 \AA, one ion was randomly rotated around the geometrical center of the other and then displaced radially to remove overlaps within a 2.5 \AA\ radius. Each random conformation was subsequently processed through a three-step optimization protocol. First, a pre-optimization was performed with the Universal Force Field (UFF) \cite{doi:10.1021/ja00051a040} using the ForceField engine of AMS. This step removes severe atomic overlaps and improves the stability of the subsequent quantum optimizations. Second, the structures were reoptimized at the GFN1-xTB level using the standard parameterization of Grimme \textit{et al.} \cite{doi:10.1021/acs.jctc.7b00118} as implemented in the DFTB engine of AMS, again including D3(BJ) corrections. Third, the resulting structures were further reoptimized at the PBE-D3(BJ) level using the ADF engine, the large frozen-core approximation, and the FIRE optimizer \cite{PhysRevLett.97.170201}.

This workflow yields three geometries for each initial ion-pair arrangement, corresponding to the UFF, GFN1-xTB, and PBE-D3(BJ) stages. The resulting collection therefore contains both equilibrium and nonequilibrium conformations and provides a broad sampling of local regions of the potential-energy surface. Since ten initial arrangements were generated for each ion pair, the final dataset contains 30 geometries per anion-cation combination. The full dataset is publicly available under the CC0 License \cite{flavian_stavarache_2026_20539350}.

\subsection{Reference calculations and property extraction}

Single-point PBE-D3(BJ) calculations were performed on all geometries in the dataset using the ADF engine of AMS. Default AMS settings were used unless otherwise noted, with the large frozen-core approximation retained and energy gradients explicitly requested. From these calculations, the reference quantities used for parameterization were extracted, namely binding energies, atomic forces, HOMO-LUMO gaps, and optimized geometries.

Binding energies were computed by subtracting the energies of the independently optimized ions from the total energy of the ion pair. Negative values therefore correspond to stable ion-pair conformations. These binding energies are approximate because the basis set superposition error (BSSE) was not corrected. This limitation should be kept in mind when interpreting absolute values, but the binding energy still provides a practical relative descriptor for comparing conformations and training the xTB models. Since total energies obtained from GFN1-xTB and PBE-D3(BJ) are not directly comparable, the use of binding energies also provides a consistent way to express relative stabilization.

Atomic forces were obtained from the negative energy gradients. Because storing the full force information for all atoms and all conformations would substantially increase the size of the training set, only the forces on selected atomic species were included, depending on the parameterization considered. The HOMO-LUMO gap was extracted directly from each single-point calculation.

\subsection{GFN1-xTB parameterization}

GFN1-xTB was parameterized using the ParAMS module of AMS \cite{doi:10.1021/acs.jcim.1c00333}. The corresponding input files were generated with the \textit{params} Python package, which was used to assemble the job collection, the training set, the parameter interface, and the engine definitions.

All parameterizations were performed with the covariance matrix adaptation evolution strategy (CMA-ES) optimizer \cite{542381}, which is well suited to the highly nonlinear and non-smooth error landscape encountered in semiempirical parameter fitting. The optimization target was the root-mean-squared error (RMSE), chosen for its more direct interpretability relative to the default sum of squared errors. Due to the size of the training sets, the data was processed in batches of 256 jobs that were randomly shuffled between optimization cycles. The population size per generation was set equal to the number of available cores.

Training runs were stopped when no progress was observed for 1000 iterations or when the wall-clock time exceeded 60 h. Benchmarking of a given parameter set was performed with the ParAMS single-point task, which evaluates all jobs in the corresponding collection and compares the calculated values with the reference data. Additional parameterization-specific settings are provided in the Supporting Information.

\begin{figure*}[t!]
    \centering
    \includegraphics[width=0.8\linewidth]{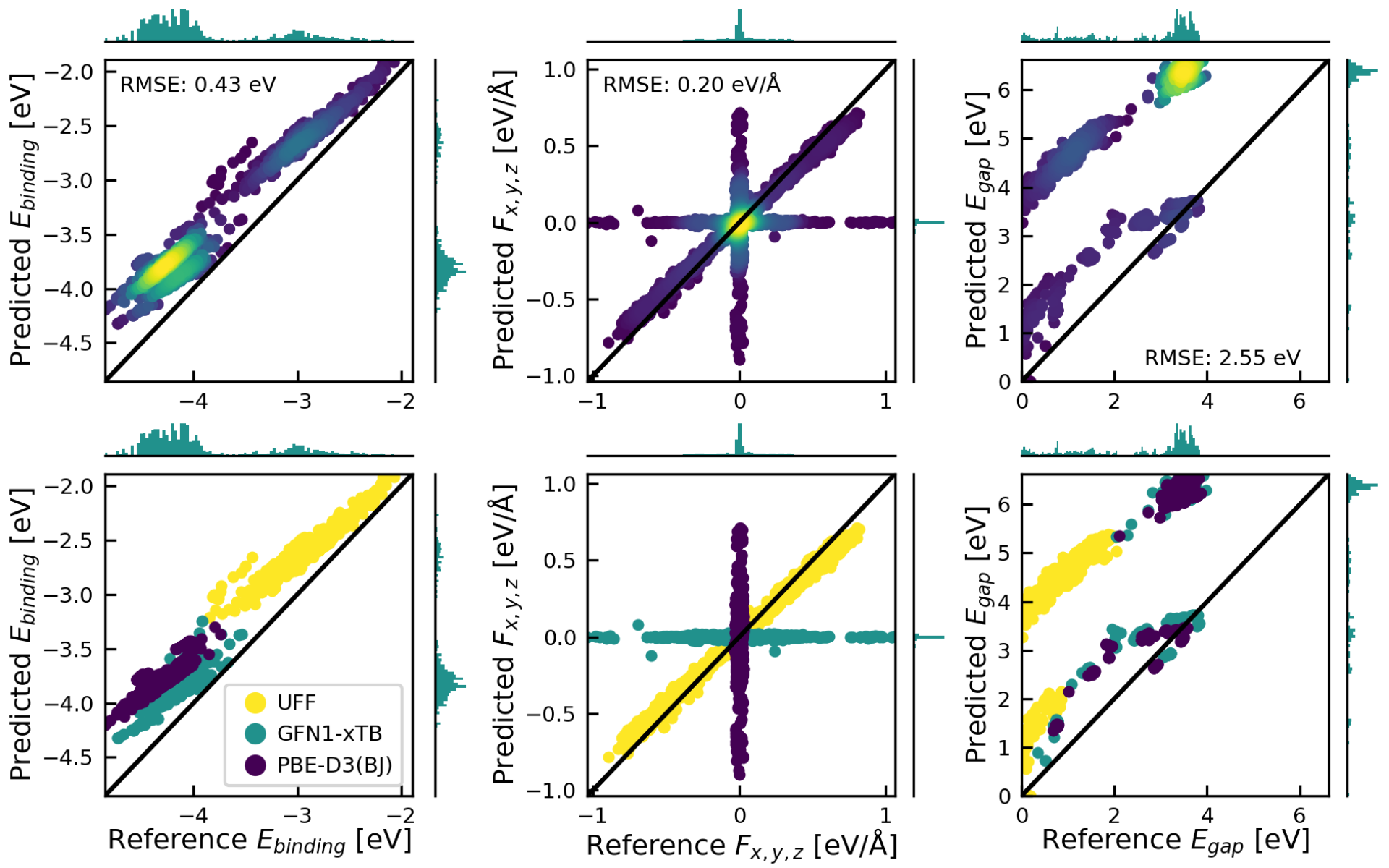}
    \caption{Dataset performance of GFN1-xTB. Columns from left to right depict binding energies, force components on the Cl anion, and HOMO-LUMO gap. On the first row, areas of lower point densities are depicted in dark blue, and higher densities in bright green to yellow. On the second row, the points are colored based on the optimization of the system.}
    \label{fig:Fig_03}
\end{figure*}

\section{Results and Discussion}
\label{sec:results}
%%%

Before introducing any parameterization strategy, the performance of the original GFN1-xTB method was evaluated on the dataset of ionic pairs containing Cl anions. This analysis establishes the strengths and limitations of the method when applied to ionic liquids and identifies systematic trends that can guide subsequent parameterization efforts. Figure~\ref{fig:Fig_03} shows parity plots comparing GFN1-xTB predictions against PBE-D3(BJ) reference values for binding energies, atomic forces, and HOMO-LUMO gaps. The dataset includes geometries obtained from UFF, GFN1-xTB, and PBE-D3(BJ) optimizations, thereby sampling both equilibrium and nonequilibrium configurations. For the binding energies, all systems lie within a relatively narrow range (approximately $-5$ to $-2$~eV), indicating that the sampled configurations correspond predominantly to stable ion pairs. GFN1-xTB captures the general trend across the dataset but exhibits systematic deviations from the reference values. In particular, the method tends to overestimate binding energies, leading to an underestimation of the stability of ionic pairs. The deviations depend on the origin of the geometries: structures optimized at the GFN1-xTB and PBE-D3(BJ) levels generally exhibit lower binding energies than those obtained from UFF pre-optimization, reflecting the absence of explicit electrostatic interactions in the latter. Despite this, no clear preference is observed in the PBE-D3(BJ) reference data, whereas GFN1-xTB shows a noticeable bias toward its own optimized geometries. The overall root-mean-squared error (RMSE) for the binding energies is ~0.43 eV.

\begin{figure*}[t!]
    \centering
    \includegraphics[width=0.8\linewidth]{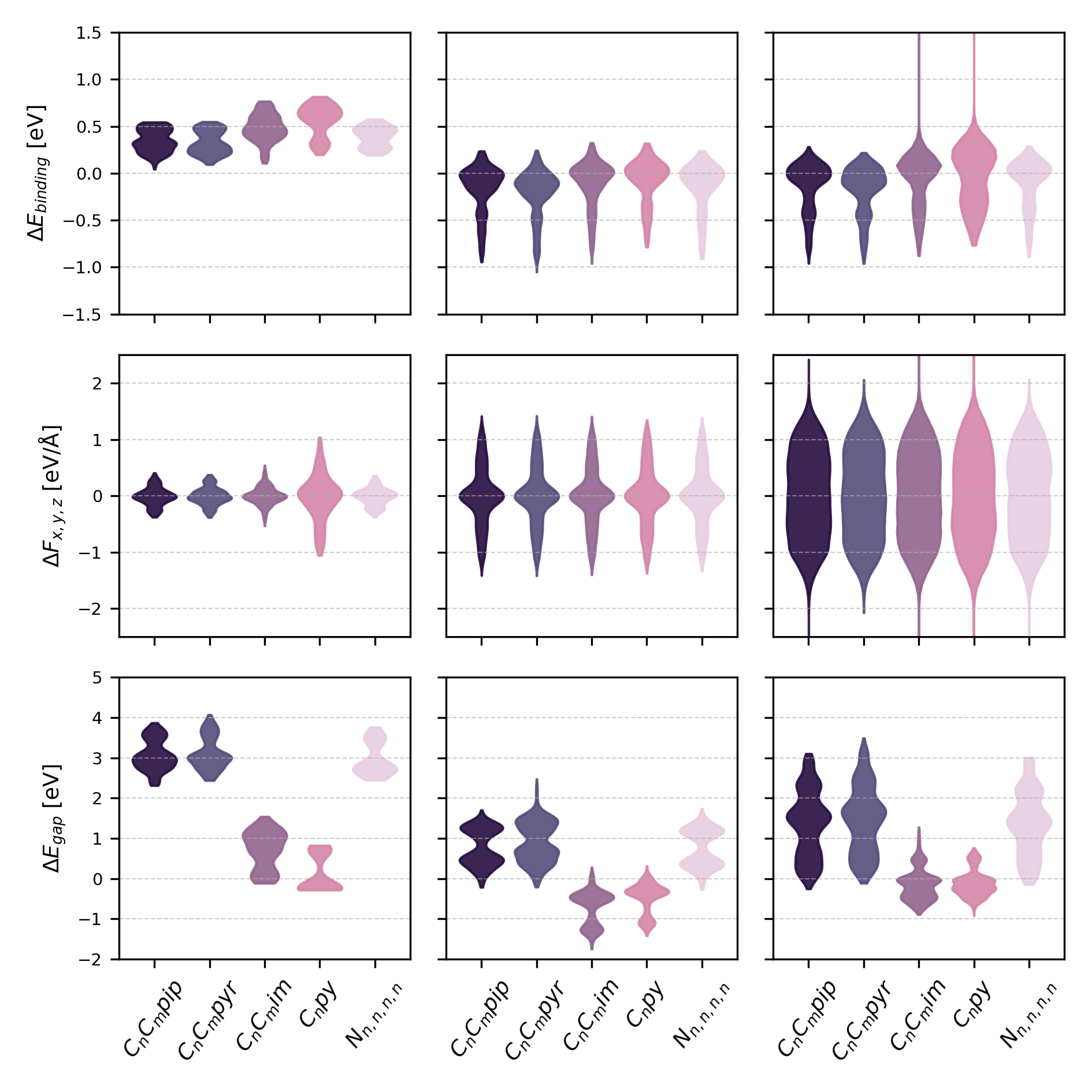}
    \caption{Distribution of errors in GFN1-xTB computed binding energies, forces, and HOMO-LUMO gaps compared to PBE-D3(BJ). The errors are split per cation species. The three columns show the results for the chloride, fluorinated compounds, and oxygen containing compounds respectively.}
    \label{fig:Fig_04}
\end{figure*}

The force predictions reveal a different type of discrepancy. As shown in the middle panel of Figure~\ref{fig:Fig_03}, a significant fraction of the data points cluster near zero force when evaluated with GFN1-xTB, particularly for structures optimized at the GFN1-xTB level. In contrast, PBE-D3(BJ) predicts non-negligible forces for many of these configurations, indicating that the equilibrium positions differ between the two methods. This effect is especially pronounced for chloride-containing systems, where the position of the anion relative to the cation is highly sensitive to the balance between electrostatic and short-range interactions. As a result, GFN1-xTB tends to predict shallow energy gradients in regions where the reference method still exhibits significant forces. The RMSE for the forces is 0.20~eV/\AA, reflecting not only quantitative differences but also qualitative discrepancies in the location of energy minima. The largest deviations are observed for the HOMO-LUMO gaps. Two distinct bands can be identified in the parity plot, one of which follows the identity line closely while the other is shifted by several electronvolts toward higher energies. Unlike the case of the forces, this separation does not correlate with the origin of the geometries but instead depends on the specific ionic systems. UFF-optimized structures tend to exhibit smaller gaps, consistent with their nonequilibrium character, but the dominant feature is the presence of two clearly separated regimes. The general RMSE of ~2.5 eV confirms that the default parameterization of GFN1-xTB is unable to provide a reliable description of the electronic levels of ionic liquids.

To further elucidate the origin of these errors, we analyzed the deviations as a function of cation and anion type. Figure~\ref{fig:Fig_04} presents the error distributions for binding energies, forces, and HOMO-LUMO gaps, grouped by cation family and by three categories of anions: chloride, fluorinated anions, and oxygen-containing anions. This classification reflects similarities in the electronic structure of the anions and their expected contributions to the molecular orbitals of the ionic pairs. For the binding energies, the error distributions show a moderate dependence on the anion type. Fluorinated and oxygen-containing anions show distributions close to zero, with asymmetric tails that indicate a tendency of GFN1-xTB to underestimate the magnitude of the binding energies. In contrast, chloride-containing systems exhibit a shift toward positive errors, consistent with the overestimation observed in the global analysis. Variations across cation families are less pronounced, although aromatic systems such as imidazolium and pyridinium show slightly larger deviations in some cases.

\begin{figure*}[t!]
    \centering
    \includegraphics[width=0.8\linewidth]{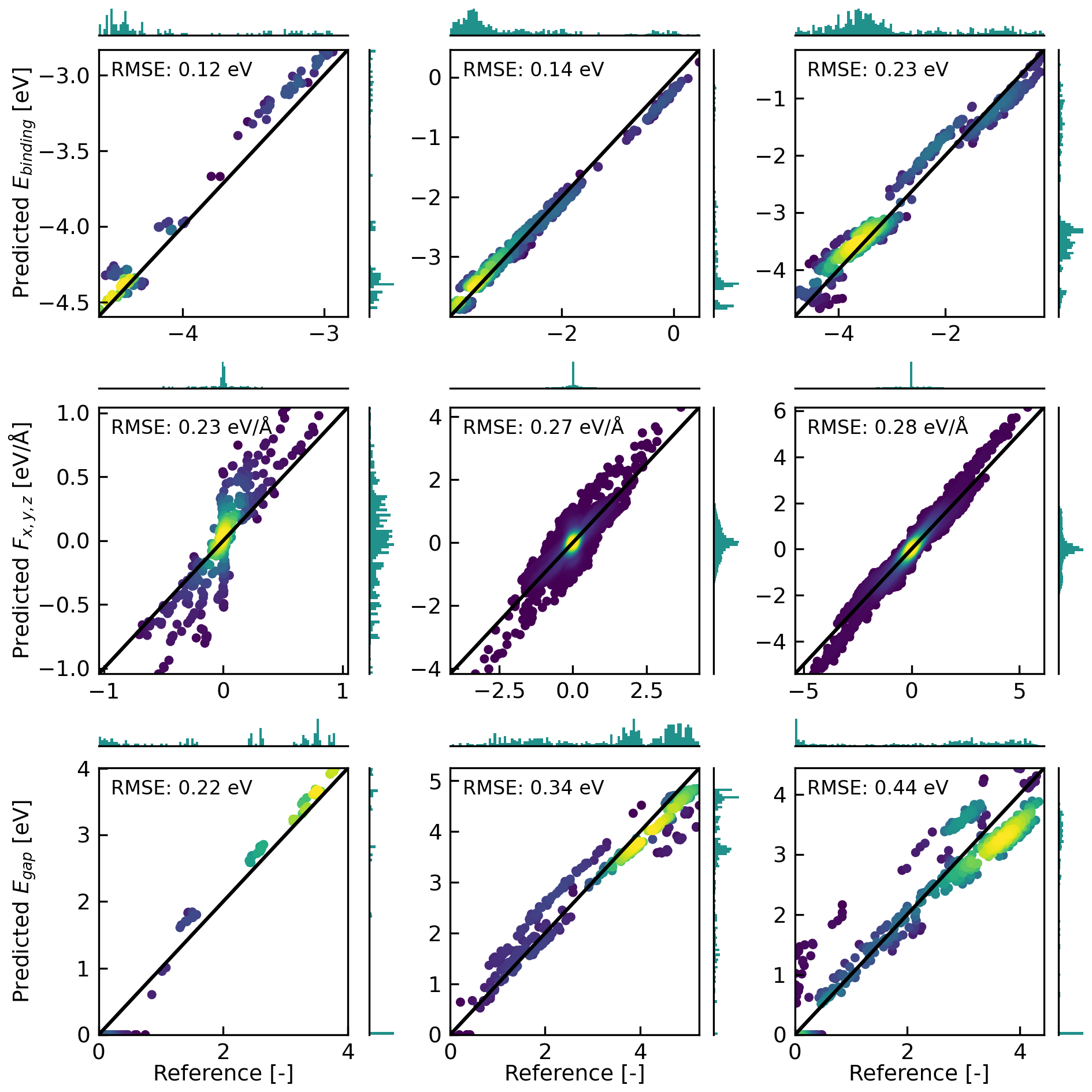}
    \caption{Dataset performance of CXMIM-xTB (GFN1-xTB trained on ionic liquids containing imidazolium-based cations). The parameterizations were trained against binding energies, forces, HOMO-LUMO gaps, and geometries. The three columns represent the results for chloride, fluorinated anions, and oxygen-containing anions from left to right.}
    \label{fig:Fig_05}
\end{figure*}

Force errors are generally centered around zero and exhibit approximately symmetric distributions, indicating that GFN1-xTB captures the overall magnitude of the forces. However, broader distributions are observed for specific combinations, most notably for pyridinium cations interacting with chloride anions, suggesting that certain local interactions are not well described. Oxygen-containing anions also show wider distributions, which points to inaccuracies in the description of interactions involving oxygen atoms. The most distinctive trends are observed for the HOMO-LUMO gaps. The error distributions separate clearly into two groups of cations: saturated systems (ammonium, pyrrolidinium, and piperidinium) and aromatic systems (imidazolium and pyridinium). This separation is consistent across all anion categories and indicates that GFN1-xTB struggles to describe both types of electronic structure within a single parameter set. The larger errors observed for saturated systems suggest that the balance between localized and delocalized electronic states is not captured consistently.

Overall, this analysis shows that while GFN1-xTB is able to reproduce general energetic trends in ionic liquids, it exhibits systematic and chemically dependent errors, particularly for electronic properties. These deficiencies are not random but follow identifiable patterns across cation families and anion types. This observation provides a basis for defining targeted parameterization strategies, either by focusing on specific classes of ionic liquids or by attempting to improve the description of both saturated and aromatic systems within a unified framework.

The trends identified in the previous section suggest that a single parameter set is unable to simultaneously describe the electronic structures of saturated and aromatic cations. To isolate and address this limitation, a targeted parameterization was performed for a subset of systems containing imidazolium-based cations. These systems provide a suitable starting point, as they represent one of the most widely studied classes of ionic liquids and exhibit a consistent electronic structure characterized by delocalized $\pi$-orbitals. To maintain a controlled level of chemical complexity, the training dataset was restricted to cations of the type \ch{[XMeim]+} (X=Me,Et,Pr,Bu,Pe,He), in which a methyl group is attached to one of the nitrogen atoms. This selection results in a set of six cations, which, when combined with the considered anions, yields 1196 ionic configuration. The dataset was further divided into three subsets based on the previously introduced anion classification: chloride (141 configurations), fluorinated anions (456 configurations), and oxygen-containing anions (599 configurations). Independent parameterizations were performed for each subset using the same training targets (binding energies, forces, HOMO-LUMO gaps, and geometries) and the same weighting scheme.

The performance of the resulting parameter sets, collectively referred to as CXMIM-xTB, is shown in Figure~\ref{fig:Fig_05}. For reference, Figure~\ref{fig:Fig_S3} shows the performance of the GFN1-xTB model on the same training data set, and Figure~\ref{fig:Fig_S4} compares the error analysis between GFN1-xTB and CXMIM-xTB, and Table ~\ref{tab:rmse} shows the respective comparison of RMSE. For chloride-containing systems, a clear improvement is observed in the prediction of electronic properties. The HOMO-LUMO gaps more closely follow the reference values, resulting in a substantial reduction in overall deviation. Binding energies are also reproduced with significantly improved accuracy, indicating that the balance between electrostatic and short-range interactions is better captured within this restricted chemical space. While the force predictions on the chloride atom show a broader spread compared to the original parameterization, the separation previously observed between different geometry classes becomes less pronounced, suggesting a more consistent description of the potential energy surface. In addition, the optimized geometries exhibit a marked reduction in outliers, as reflected in the narrower distribution of RMSE values (Table ~\ref{tab:rmse}).

For fluorinated and oxygen-containing anions, improvements are observed primarily in the energetic and structural properties. Binding energies show a more consistent agreement with the reference data, and the forces are described with significantly higher accuracy, particularly for oxygen-containing systems. The reduction in force errors is especially relevant for applications involving structural relaxation or molecular dynamics. In contrast to the chloride and fluorinated anions, the HOMO-LUMO gaps for the oxygen-containing systems remain largely unchanged, indicating that the dominant source of error for these anions is not fully addressed by modifying the oxygen atomic parameters alone. This behavior suggests that the electronic structure of the cation plays a key role in determining the frontier orbital energies of the ionic pairs. 

Across all anion categories, the structural predictions benefit from the targeted parameterization. The spread of RMSE values decreases noticeably, indicating a reduction in poorly described geometries, while the mean deviations remain comparable to the original model. This reflects a more robust description of the local energy landscape without introducing significant distortions in the equilibrium structures. Overall, the CXMIM-xTB parameterization demonstrates that focusing on a well-defined subset of ionic liquids leads to a more accurate and consistent description of both energetic and electronic properties. Although the resulting model remains specialized to imidazolium-based systems, it provides a substantially improved starting point for studying this important class of ionic liquids and highlights the potential of targeted parameterization strategies within the xTB framework.

The improved description of electronic properties provided by CXMIM-xTB allows its application to quantities of direct relevance for electrochemical systems. In particular, ionic liquids are widely used as electrolytes due to their chemical stability and low volatility, and their performance is largely determined by the accessible electrochemical window (ECW). The ECW defines the potential range within which the electrolyte remains stable and is therefore a key parameter for the design of battery and capacitor systems. Within a quantum mechanical framework, the ECW can be approximated from the HOMO-LUMO gap of the system, which provides an upper bound for the voltage range in which no electronic breakdown occurs. A commonly used approach to estimate this quantity combines molecular dynamics simulations with electronic-structure calculations. In the work of Zhang \textit{et al.} \cite{doi:10.1021/jp5034257}, inspired by Ong \textit{et al.} \cite{doi:10.1021/cm200679y}, ionic-liquid systems are first equilibrated using classical molecular dynamics, followed by the extraction of representative snapshots. These configurations are then either relaxed at the DFT level (AIMD-min) or directly evaluated using single-point calculations (AIMD-sp), typically employing hybrid functionals such as B3LYP with extended basis sets. While this multiscale strategy provides reliable ECW estimates, it remains computationally demanding, particularly when several ion pairs are required to represent the liquid environment.

\begin{figure} [t!]
    \centering
    \includegraphics[width=0.9\linewidth]{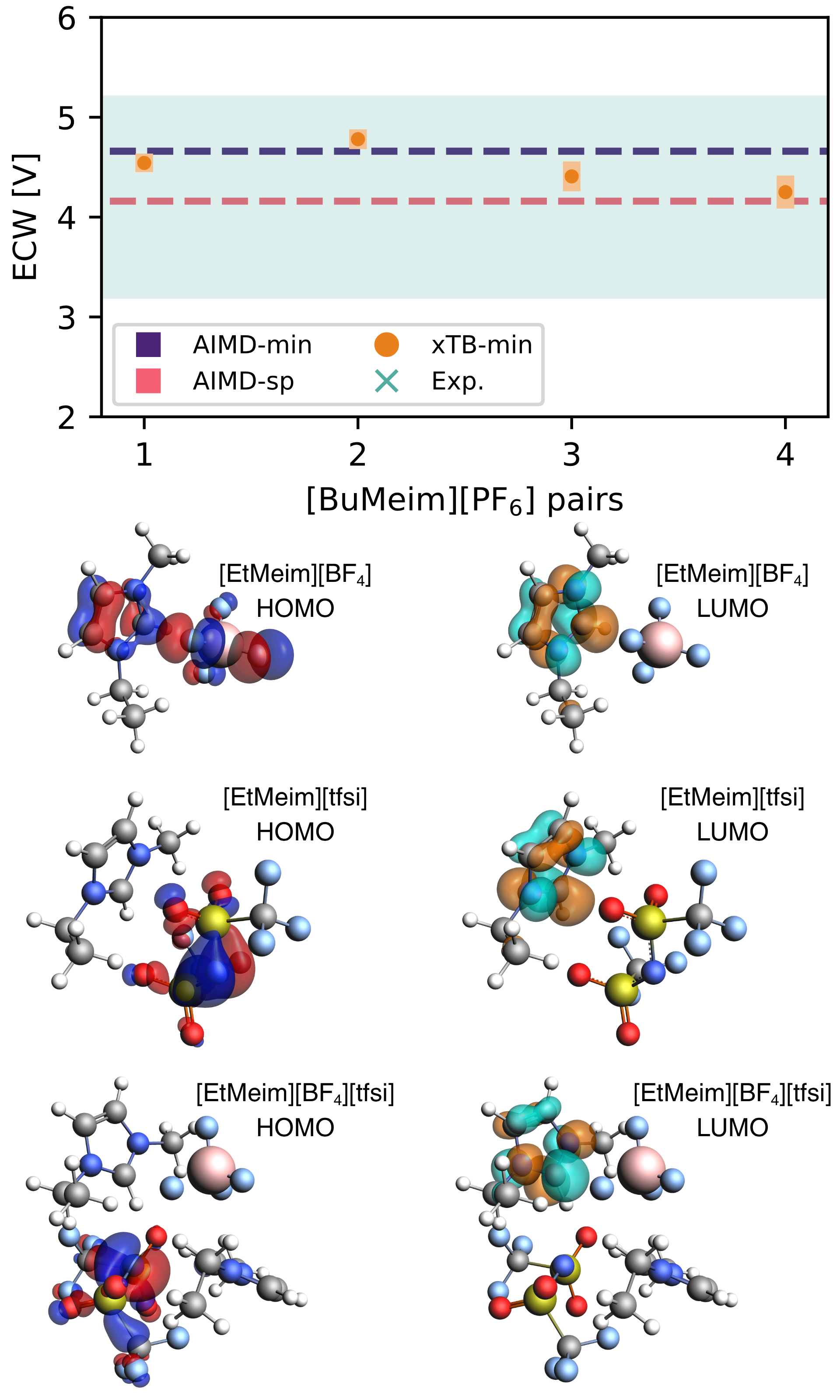}
    \caption{\ch{[BuMeim][PF6]} ECW dependency on the number of ionic pairs compared to ab-inito calculations \cite{doi:10.1021/jp5034257} and experimental data \cite{B007172M, doi:10.1021/je800678e, Lewandowski2010, BARISCI200422, B305734H} (top). Randomly generating ionic clusters and optimizing them with CXMIM-xTB results in similar ECWs. Molecular orbitals of \ch{[EtMeim][BF4]}, \ch{[EtMeim][tfsi]}, and \ch{[EtMeim][BF4][tfsi]} (bottom).}
    \label{fig:Fig_06}
\end{figure}

The CXMIM-xTB model provides an alternative route to estimate ECWs without relying on molecular dynamics simulations. In the present approach, random ionic configurations are generated following the same procedure used for dataset construction and subsequently optimized using CXMIM-xTB. For each system, 100 random conformations are generated, and the HOMO-LUMO gap is averaged over the 20 lowest energy structures. This procedure effectively replaces the sampling obtained from molecular dynamics with a direct exploration of the configurational space through geometry optimization, significantly reducing the computational cost while retaining access to representative low-energy structures. The dependence of the ECW on the number of ion pairs was first assessed using \ch{[BuMeim][PF6]} systems containing up to four ion pairs. The resulting ECW values show only minor variations across system sizes, remaining within the range of approximately 4.25-4.78~V (Figure~\ref{fig:Fig_06}). This behavior indicates that even single ion-pair calculations can provide a reasonable approximation of the electrochemical window for these systems, with limited benefit from increasing the number of ions. At the same time, the computational cost increases significantly with system size, suggesting that small clusters already capture the dominant electronic features relevant for ECW estimation.

A broader comparison is presented in Table ~\ref{tab:ecw-results}, where ECWs obtained using CXMIM-xTB for one and two ion pairs are compared with literature values obtained from AIMD-min and AIMD-sp approaches \cite{doi:10.1021/jp5034257}, as well as available experimental data. For ionic liquids containing fluorinated anions, such as \ch{[BuMeim][PF6]} and \ch{[BuMeim][BF4]}, the CXMIM-xTB results are in close agreement with both computational and experimental references. For systems involving the more flexible \ch{[tfsi]-} anion, single ion-pair calculations underestimate the ECW, reflecting the sensitivity of the electronic structure to conformational variations of the anion. This effect is mitigated when two ion pairs are considered, leading to ECW values that are consistent with the AIMD-based estimates and fall within the experimental range. Overall, these results indicate that the CXMIM-xTB approach provides an efficient and reliable alternative for estimating electrochemical windows, particularly when appropriate configurational sampling is ensured.

\begin{table*} [t!]
    \centering
    \caption{Comparison between the ECW computed using CXMIM-xTB on one and two ionic pairs (this work), AIMD-min \cite{doi:10.1021/jp5034257}, AIMD-sp \cite{doi:10.1021/jp5034257}, and experimental data \cite{B007172M, doi:10.1021/je800678e, Lewandowski2010, BARISCI200422, B305734H} for \ch{[BuMeim][PF6]}, \ch{[BuMeim][BF4]}, \ch{[BuMeim][tfsi]}, and \ch{[EtMeim][tfsi]}. The values obtained in this work are similar to the ones obtained using AIMD-min and are within experimental ranges.}
    \begin{tabular}{cccccc}
        Ionic & CXMIM-xTB (tw) & CXMIM-xTB (tw) & AIMD-min & AIMD-sp & Exp. \\
        Liquid & x1 ECW [V] & x2 ECW [V] & ECW [V] & ECW [V] & ECW [V] \\
        \hline
        \hline
        \ch{[BuMeim][PF6]} & 4.54 $\pm$ 0.09 & 4.78 $\pm$ 0.10 & 4.66 $\pm$ 0.03 & 4.16 $\pm$ 0.02 & 3.2-5.2 \\
        \ch{[BuMeim][BF4]} & 4.62 $\pm$ 0.07 & 4.62 $\pm$ 0.13 & 4.49 $\pm$ 0.05 & 4.05 $\pm$ 0.06 & 4.1-4.6 \\
        \ch{[BuMeim][tfsi]} & 3.55 $\pm$ 0.06 & 4.15 $\pm$ 0.15 & 4.19 $\pm$ 0.05 & 2.35 $\pm$ 0.10 & 4.3-4.6 \\
        \ch{[EtMeim][tfsi]} & 3.51 $\pm$ 0.10 & 4.09 $\pm$ 0.13 & 4.30 $\pm$ 0.05 & - & 4.0-5.0 \\
    \end{tabular}
    \label{tab:ecw-results}
\end{table*}

Beyond global properties such as the ECW, the CXMIM-xTB model can also be used to analyze the electronic structure of pure and mixed ionic liquids. To illustrate this, four representative systems were considered: the ionic pairs \ch{[EtMeim][BF4]} and \ch{[EtMeim][tfsi]}, a charged trimer \ch{[EtMeim][BF4] + [tfsi]-}, and the neutral tetramer \ch{[EtMeim][BF4][tfsi]}. These systems provide insight into the behavior of both pure ionic liquids and binary mixtures. For each case, 100 random conformations were generated and optimized using CXMIM-xTB, and the lowest-energy structure was selected for analysis (see Figure~\ref{fig:Fig_S5}). The distribution of the frontier molecular orbitals for these systems is also shown in Figure~\ref{fig:Fig_06}. In all cases, the LUMO is consistently localized on the imidazolium cation, indicating that the reduction process is dominated by the cationic species. In contrast, the HOMO exhibits a more variable character. For \ch{[EtMeim][BF4]}, the HOMO is delocalized across both ions, while in systems containing \ch{[tfsi]-} it is predominantly localized on the anion. In the mixed system, the HOMO is found exclusively on the \ch{[tfsi]-} anion, suggesting that this species is more electrochemically active than \ch{[BF4]-}. These trends are consistent with previous studies based on hybrid DFT calculations \cite{TRENZADO2021116049}, indicating that CXMIM-xTB is able to reproduce the key features of the electronic structure in both pure and mixed ionic liquids. This further supports the applicability of the method for the analysis of ionic interactions and electronic properties in complex electrolyte systems.

In addition to electrochemical properties, the optical response of ionic liquids is of growing interest for applications in photonics, sensing, and optoelectronic devices. In this context, the UV-Vis absorption spectrum provides direct insight into the electronic excitations of the system and their relation to the underlying molecular structure. Given the improved description of ground-state properties obtained with CXMIM-xTB, it is therefore natural to assess its performance for excited-state calculations.

\begin{figure} [t!]
    \centering
    \includegraphics[width=0.9\linewidth]{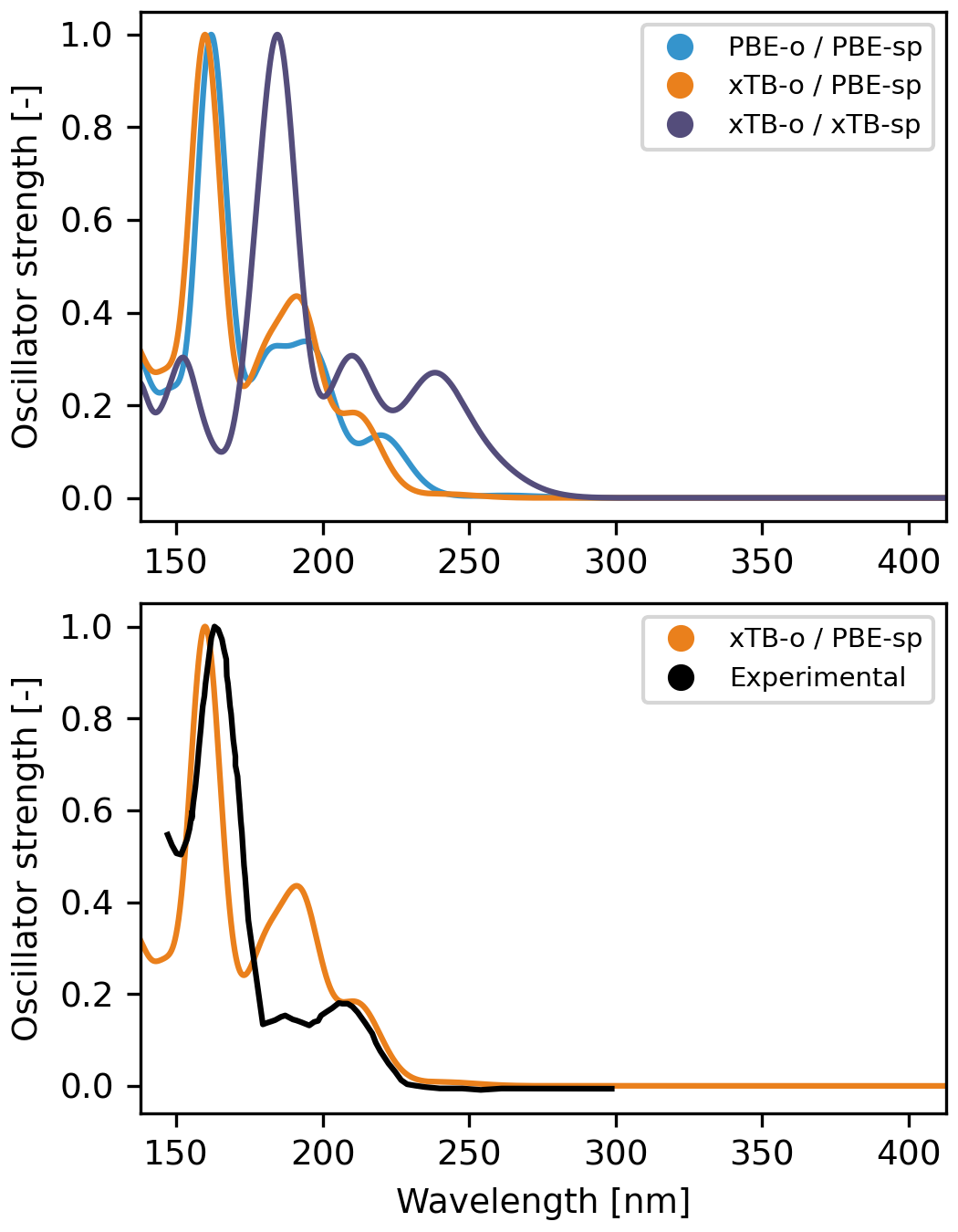}
    \caption{Comparison of simulated and experimental \cite{C6CP02930B} UV/Vis spectra for the \ch{[BuMeim][BF4]} IL ion pairs. The simulated spectra were calculated using a combination of DFT and DFTB, where -o and -sp in the legend stand for geometry optimization and single point calculations, respectively.}
    \label{fig:Fig_07}
\end{figure}

Figure~\ref{fig:Fig_07} compares the UV-Vis spectra of \ch{[BuMeim][BF4]} obtained using time-dependent DFTB (TD-DFTB) based on the CXMIM-xTB parameterization, as well as reference spectra computed using sTD-DFT at the level of PBE-D3 (BJ) and available experimental data \cite{C6CP02930B}. Although the CXMIM-xTB model provides an improved description of the HOMO-LUMO gap, the resulting TD-DFTB spectra remain qualitatively similar to those obtained with the original GFN1-xTB parameters. In particular, the position of the main absorption peak exhibits a systematic shift relative to the DFT reference and experimental results. This behavior indicates that the remaining discrepancies in the UV-Vis spectra are not only determined by the accuracy of the frontier orbital energies but also depend on the description of excited states, which is not explicitly targeted in the present parameterization. As a result, the direct use of CXMIM-xTB within a TD-DFTB framework does not fully resolve the limitations observed for optical properties of ionic liquids. Despite this, the improved structural description provided by CXMIM-xTB can be used within a hybrid computational strategy. As shown in Figure~\ref{fig:Fig_07}, performing sTD-DFT calculations on optimized geometries with CXMIM-xTB produces spectra that are nearly indistinguishable from those obtained using fully DFT-optimized geometries. This result suggests that the main source of error in the TD-DFTB spectra lies in the excited-state treatment rather than in the underlying geometries.

This observation opens the possibility of replacing the geometry optimization step in conventional workflows with CXMIM-xTB calculations, while retaining DFT-based methods for the evaluation of excited states. In practice, this approach enables the generation of a large number of low-energy conformations at a significantly reduced computational cost, which can then be used for ensemble averaging of UV-Vis spectra. Given that geometry optimizations with xTB methods are several orders of magnitude faster than their DFT counterparts, this strategy allows for more extensive sampling of configurational space and the use of more advanced exchange-correlation functionals in the subsequent TD-DFT calculations.

For the \ch{[BuMeim][BF4]} system, the combined CXMIM-xTB geometry optimization and sTD-DFT spectral calculation reproduces the experimental spectrum with good accuracy, in line with previous studies \cite{C6CP02930B}. These results highlight that, while CXMIM-xTB does not directly improve the description of excited states within TD-DFTB, it provides an efficient and reliable foundation for hybrid workflows aimed at predicting the optical properties of ionic liquids.

The results obtained with CXMIM-xTB demonstrate that targeted parameterization can substantially improve the description of ionic liquids when the chemical space is restricted. However, the analysis of the baseline GFN1-xTB performance (Figure~\ref{fig:Fig_04}) also revealed that the largest discrepancies arise from the simultaneous description of saturated and aromatic cations. In particular, saturated systems exhibit significantly larger errors in the HOMO-LUMO gap, especially in combination with chloride anions, indicating that a more general parameterization requires a different training strategy that explicitly accounts for these distinct electronic environments. To address this limitation, a hierarchical parameterization approach was adopted. Rather than fitting all parameters simultaneously on the full dataset, the training was divided into sequential steps targeting specific atomic species and chemical environments. In the first step, the parameterization was restricted to saturated cation systems, where GFN1-xTB already provides a relatively consistent description. Within this subset, the parameters associated with the chlorine atom were optimized in order to improve the representation of its atomic orbitals and their contribution to the electronic structure of the ionic pairs. This step primarily targets the large deviations observed for chloride-containing systems.

\begin{figure*} [t!]
    \centering
    \includegraphics[width=0.8\linewidth]{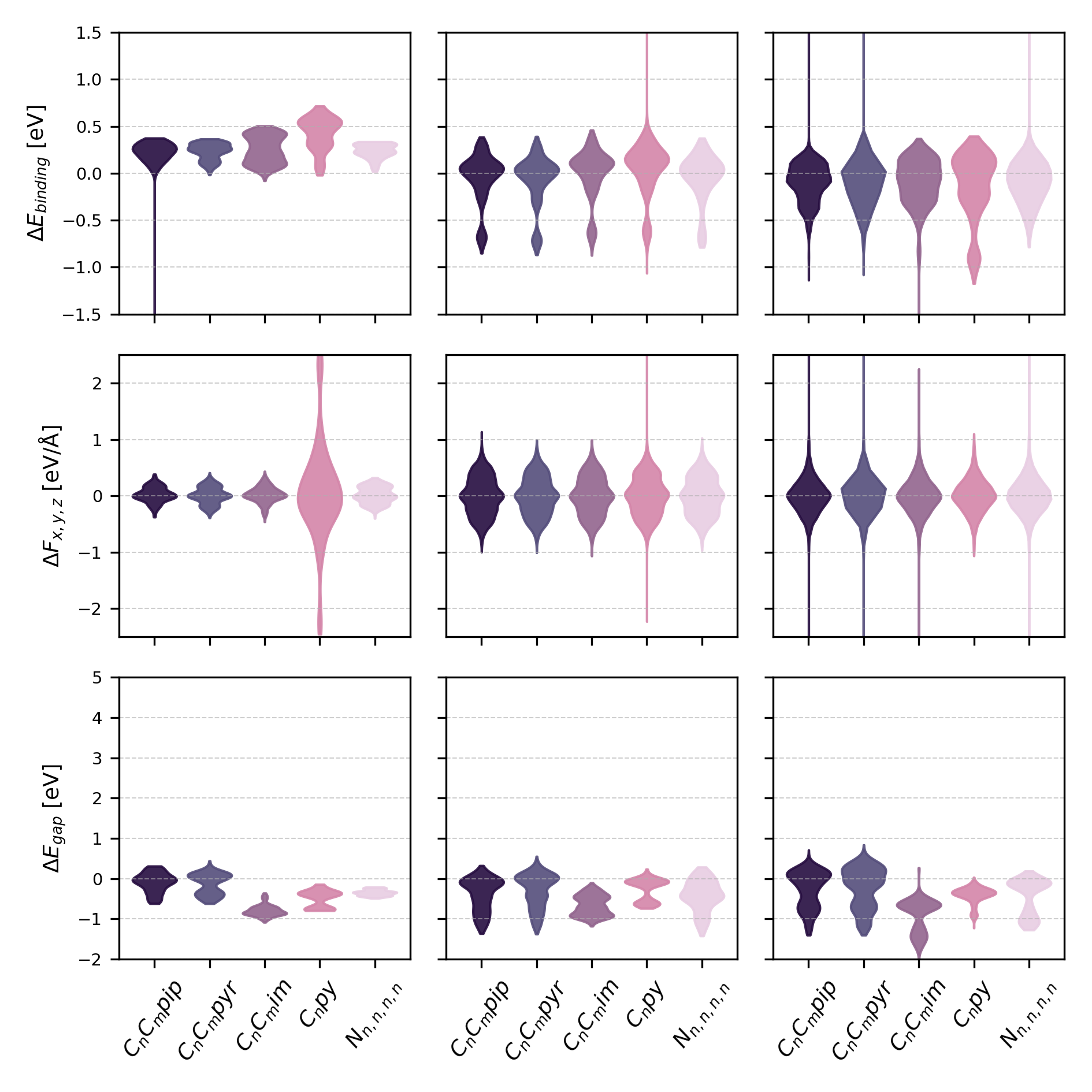}
    \caption{Distribution of errors in IL-xTB computed binding energies, forces, and HOMO-LUMO gaps compared to PBE-D3(BJ). The errors are split per cation species. The three columns show the results for the chloride, fluorinated compounds, and oxygen-containing compounds, respectively.}
    \label{fig:Fig_08}
\end{figure*}

In a second step, the chlorine parameters were fixed and the training dataset was extended to include both saturated and aromatic cations. The parameters associated with carbon were then optimized, focusing on the $2s$ and $2p$ orbitals. By including both cation types at this stage, we aim to obtain carbon parameters that consistently describe their different electronic environments. The same strategy was subsequently applied to fluorinated and oxygen-containing anions, where the relevant $s$- and $p$-subshell parameters, together with the $\alpha$ and $Z$ terms, were refined using datasets composed of saturated systems. The resulting parameter set is denoted as IL-xTB, and the full list of optimized parameters is reported in Table~\ref{tab:il_parameters}.

\begin{figure*} [t!]
    \centering
    \includegraphics[width=0.9\linewidth]{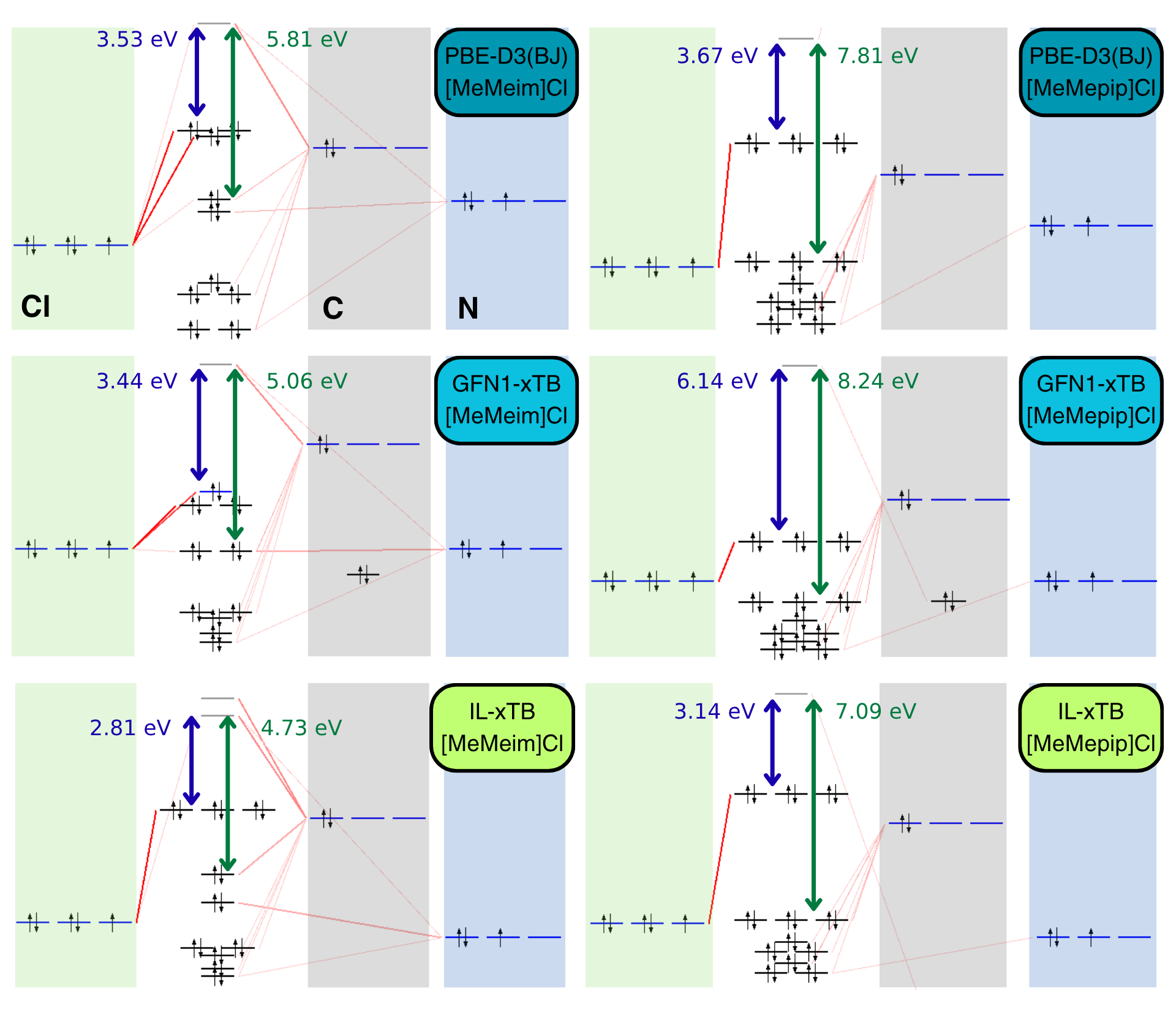}
    \caption{Atomic contributions to the molecular orbitals for \ch{[MeMeim]Cl} and \ch{[MeMepip]Cl} resulting from GFN1-xTB, IL-xTB and PBE-D3(BJ) simulations. Horizontal black lines show the electronic levels, while the black arrows show the level occupation. The red lines show the contribution of each electronic level to the molecular orbitals. The green, blue, and black columns depict the \ch{Cl}, \ch{N}, and \ch{C} contributions, respectively. The blue and green arrows represent the HOMO-LUMO gap and the HOMO-1-LUMO gaps, respectively.}
    \label{fig:Fig_09}
\end{figure*}

The performance of IL-xTB is summarized in Figure~\ref{fig:Fig_08} and Figure~\ref{fig:Fig_S6}, which shows the error distributions and parity plots for binding energies, atomic forces, and HOMO-LUMO gaps, grouped by cation family and anion class. Compared to the original GFN1-xTB results, a clear improvement is observed across most properties, although in some cases the force RMSE for chlorine increases. Most notably, the separation previously observed between saturated and aromatic cations in the HOMO-LUMO gap is no longer present, indicating that the new parameterization achieves a more balanced description of different electronic structures. This represents a key step toward transferability, as a single parameter set is now able to capture the trends across chemically diverse ionic liquids. For the binding energies, the distributions are generally centered closer to zero and exhibit reduced asymmetry, reflecting a more consistent description of interionic interactions. The improvements are observed across all anion classes, including chloride, fluorinated, and oxygen containing systems. Similarly, the force distributions show a reduced overall spread, indicating that the potential energy surface is described more uniformly. At the same time, a small number of outliers remain, particularly for specific combinations such as pyridinium-based cations interacting with chloride anions. These cases are associated with short interionic distances, where the modified carbon parameters can favor stronger interactions.

In the case of the HOMO-LUMO gaps, IL-xTB significantly reduces the overall error and eliminates the systematic split between different cation classes ((Table ~\ref{tab:rmse})). While a slight underestimation of the gap is observed on average, the deviation is consistent across systems, which suggests that relative trends can still be reliably captured. Overall, IL-xTB improves upon the limitations of the original GFN1-xTB and provides a more consistent description of the structural, energetic, and electronic properties across the ionic-liquid families considered in this work.

The improved performance of IL-xTB across different ionic-liquid families suggests that the new parameterization captures more consistently the underlying electronic structure. To further rationalize this behavior, the atomic contributions to the molecular orbitals were analyzed for representative systems containing both aromatic and saturated cations. Two ionic pairs were selected for this purpose: \ch{[MeMeim]Cl}, representing an aromatic imidazolium-based system, and \ch{[MeMepip]Cl}, representing a saturated piperidinium-based system. In both cases, short alkyl chains were chosen to maximize the interaction between the cationic ring and the chloride anion. The \ch{Cl-} anion was selected as a reference due to its relatively simple electronic structure, which facilitates the interpretation of orbital contributions. Single-point calculations were performed at the PBE-D3(BJ), GFN1-xTB, and IL-xTB levels on the lowest-energy DFT-optimized geometries. The resulting atomic contributions to the molecular orbitals are shown in Figure~\ref{fig:Fig_09}.

\begin{figure} [t!]
    \centering
    \includegraphics[width=0.8\linewidth]{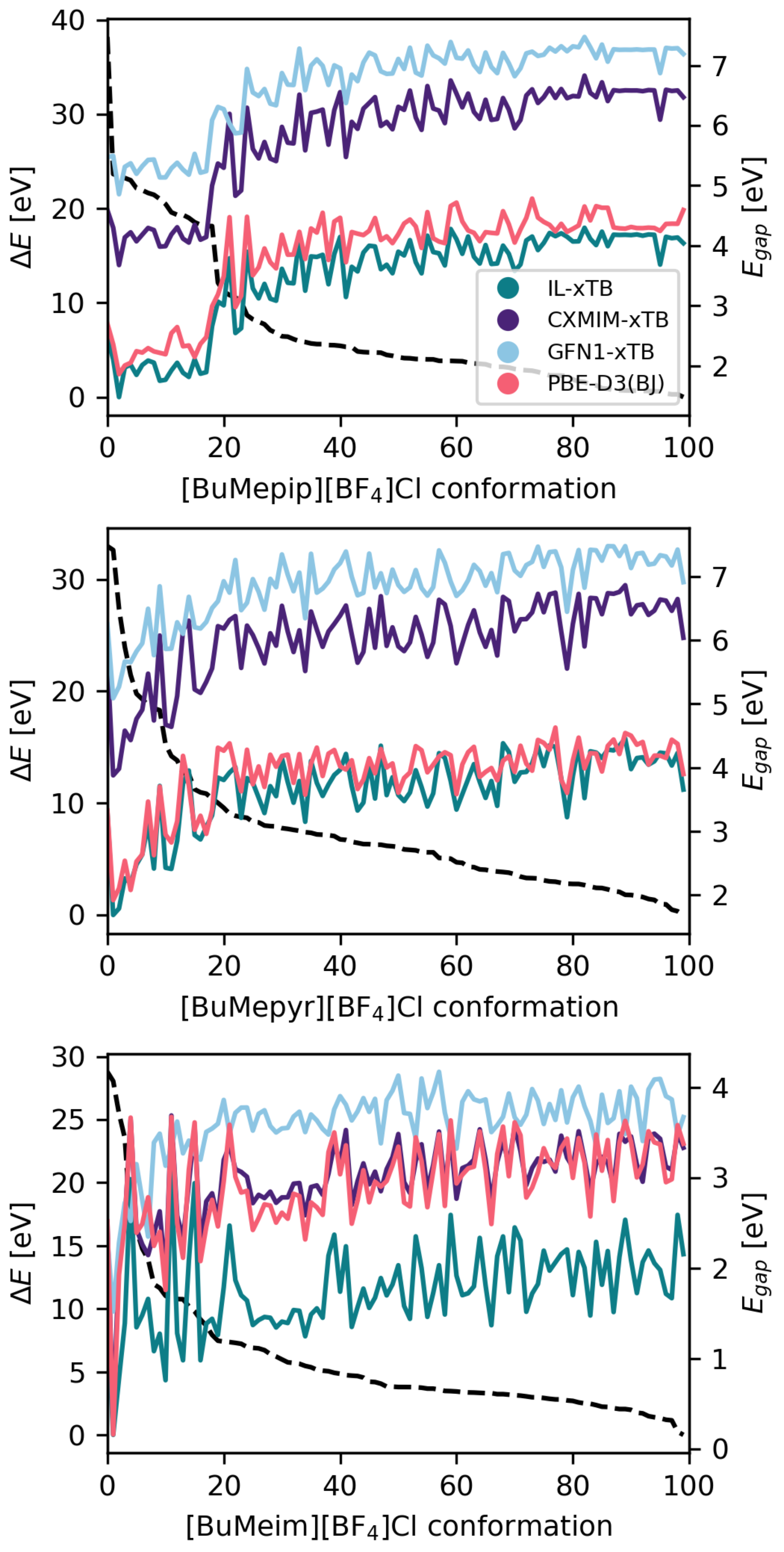}
    \caption{Comparison of HOMO-LUMO gaps calculated for \ch{[BuMepip][BF4]Cl} (top), \ch{[BuMepyr][BF4]Cl} (middle), and \ch{[BuMeim][BF4]Cl} (bottom) clusters using PBE-D3(BJ), GFN1-xTB, CXMIM-xTB, and IL-xTB. The left axis measures the relative energy of a cluster with respect to the lowest-energy cluster found. The right axis measures the HOMO-LUMO gap of the corresponding cluster.}
    \label{fig:Fig_10}
\end{figure}

For the aromatic system, \ch{[MeMeim]Cl}, the PBE-D3(BJ) results show a clear separation between the cation derived orbitals and the anion contribution, which appears as an intermediate state located between the highest occupied orbitals of the cation. In this picture, the HOMO is predominantly localized on the anion, while the LUMO resides on the cation. The GFN1-xTB model qualitatively reproduces this ordering, but exhibits noticeable discrepancies in the relative energy levels. In particular, the separation between the anion derived HOMO and the highest occupied cation orbital (HOMO$-1$) is underestimated, while the gap between the cation orbitals is also reduced. These compensating effects lead to an apparent agreement in the overall HOMO-LUMO gap for aromatic systems, consistent with the trends observed in the previous sections.

A similar analysis for the saturated system, \ch{[MeMepip]Cl}, reveals a different behavior. In this case, the discrepancies between GFN1-xTB and PBE-D3(BJ) do not compensate, resulting in a significant deviation in the HOMO-LUMO gap. While the relative spacing between the cation orbitals is reasonably reproduced, the position of the anion derived states is shifted, leading to an incorrect description of the frontier orbitals. This observation is consistent with the larger errors previously reported for saturated cations and highlights the sensitivity of the electronic structure to the balance between cation and anion contributions.

The IL-xTB parameterization provides a more consistent description across both systems. As shown in Figure~\ref{fig:Fig_09}, the relative ordering of the molecular orbitals and the contributions from the different atomic species are more closely aligned with the PBE-D3(BJ) reference. In particular, the carbon derived orbitals exhibit a more realistic energy distribution, with reduced contributions from artificially high-lying $2s$ states. At the same time, the anion derived levels are positioned more consistently relative to the cation orbitals for both aromatic and saturated systems. Although the absolute HOMO-LUMO gaps obtained with IL-xTB remain slightly underestimated, the improved alignment of the underlying orbital structure explains the more uniform error distributions observed across different cation families. This analysis shows that the hierarchical parameterization strategy improves both global properties and the underlying description of the electronic structure, providing a more balanced representation across different cation families.

Finally, we evaluated the different parameterizations across conformational ensembles that were not restricted to the optimized structures used in the previous orbital analysis. Figure~\ref{fig:Fig_10} compares the HOMO-LUMO gaps predicted by GFN1-xTB, CXMIM-xTB, IL-xTB, and PBE-D3(BJ) for 100 conformations of three representative clusters containing piperidinium, pyrrolidinium, and imidazolium cations. The conformations are ordered according to their relative PBE-D3(BJ) energies, providing a comparison across configurations ranging from high-energy structures to the most stable conformations. For the saturated \ch{[BuMepip][BF4]Cl} and \ch{[BuMepyr][BF4]Cl} systems, IL-xTB closely follows the PBE-D3(BJ) HOMO-LUMO gaps. In contrast, both GFN1-xTB and CXMIM-xTB systematically overestimate the gaps for these systems. These results confirm that hierarchical parameterization substantially improves the description of saturated cations, one of the main limitations identified for the original GFN1-xTB model.

For the aromatic \ch{[BuMeim][BF4]Cl} system, CXMIM-xTB provides the closest agreement with PBE-D3(BJ), consistent with its targeted parameterization for imidazolium-based ionic liquids. IL-xTB instead systematically underestimates the HOMO-LUMO gap, showing that extending the parameterization to a broader chemical space introduces a compromise in accuracy for this specific cation family. Nevertheless, IL-xTB considerably reduces the original GFN1-xTB dependence on cation type and provides a more balanced description across saturated and aromatic systems. Together with the error distributions and molecular-orbital analysis in Figures~\ref{fig:Fig_08} and \ref{fig:Fig_09}, these results demonstrate the potential of hierarchical parameterization to improve the description of chemically diverse ionic-liquid systems. They also highlight the complementary roles of the two models developed here: CXMIM-xTB provides a more accurate description within a restricted chemical space and can be directly applied to targeted ionic-liquid applications, whereas IL-xTB demonstrates how chemically informed hierarchical training can extend the methodology toward a broader range of ionic systems. Importantly, these improvements are achieved while retaining the substantially lower computational cost of DFTB compared with DFT (see the performance benchmarks presented in Figures S7 and S8 of the ESI).

\section{Conclusions}
\label{sec:conclusions}
%%%

In this work, we investigated reparameterizing GFN1-xTB to describe the structural, energetic, and electronic properties of ionic liquids. We constructed a database containing more than 14,000 conformations using a combination of classical, semiempirical, and DFT calculations, providing a chemically diverse dataset for training and evaluating different parameterization strategies. Our analysis of the original GFN1-xTB model revealed that its performance strongly depends on the chemical environment, with systematic deviations in binding energies and forces and particularly large differences in the description of HOMO-LUMO gaps for saturated and aromatic cations. These results highlight the importance of explicitly considering the chemical diversity of ionic systems when developing semiempirical parameterizations.

Based on this analysis, we explored two complementary parameterization strategies. First, we developed CXMIM-xTB as a targeted model for imidazolium-based ionic liquids. By restricting the chemical space, we optimized the parameters while maintaining the original GFN1-xTB framework and substantially improved the description of binding energies, geometries, and frontier orbital energies. Beyond the training properties, we applied CXMIM-xTB to predict the electrochemical windows of pure and mixed ionic liquids. The calculated values agree well with DFT-based approaches and available experimental ranges, while the reduced computational cost allows extensive sampling of low-energy ionic configurations.

We further examined the applicability of CXMIM-xTB to optical properties. Direct TD-DFTB calculations with the reparameterized model do not significantly improve the UV-Vis spectra compared with the original GFN1-xTB parameterization, despite the improved HOMO-LUMO gaps. This result indicates that improving ground-state frontier orbital energies alone does not provide an accurate description of the excited-state response. Nevertheless, TD-DFT calculations using geometries optimized with CXMIM-xTB produce essentially the same spectra as those using DFT-optimized geometries. CXMIM-xTB can therefore provide an efficient configurational sampling and geometry optimization step before higher-level excited-state calculations, substantially reducing the computational cost of such hybrid workflows.

We developed a second parameterization strategy to explore whether we could extend these improvements to a broader range of ionic liquids. For the resulting IL-xTB model, we employed a hierarchical training procedure that accounts for the different electronic environments of saturated and aromatic cations and different anion families. IL-xTB substantially reduces the systematic errors observed with the original GFN1-xTB across the broader chemical space and provides a more consistent description of the underlying molecular orbital contributions. Although further validation across additional chemical families and properties will be required before IL-xTB can be considered a generally transferable parameterization, our results provide a proof of concept that chemically informed training strategies can systematically address the limitations identified in the original model.

Overall, our results demonstrate two complementary routes for extending semiempirical tight-binding methods to complex ionic systems. Targeted parameterizations such as CXMIM-xTB can provide accurate and computationally efficient models for well-defined chemical families and specific applications, whereas IL-xTB demonstrates how hierarchical parameterization can improve the description of ionic liquids across different cation and anion families. More generally, we show that combining extensive configurational sampling, DFT reference data, and chemically informed parameter optimization provides a systematic framework for developing efficient quantum-mechanical models of ionic liquids and other chemically diverse molecular systems.\\

%%%  END OF MAIN TEXT    %%%%%%%%%%%%%%%%%%%%%%%%%%%%%%%
%%%%%%%%%%%%%%%%%%%%%%%%%%%%%%%%%%%%%%%%%%%%%%%%%%%%%%%%
%%%%%%%%%%%%%%%%%%%%%%%%%%%%%%%%%%%%%%%%%%%%%%%%%%%%%%%%
%%%%%%%%%%%%%%%%%%%%%%%%%%%%%%%%%%%%%%%%%%%%%%%%%%%%%%%%

% \section*{Supporting Information}

%Here the SI description.

%\section*{Acknowledgements}

%S.C. acknowledges funding from the Irène Curie Fellowship program of the Eindhoven University of Technology.

%\section*{Author contributions}

%Here the author contributions.

\section*{Competing interests}

The authors declare no competing interests.

\onecolumngrid

\bibliography{references}% Produces the bibliography via BibTeX.

%Uncomment to add the TOC figure at the end of the manuscript

\newpage

\section*{Table of content image}

\begin{figure}[htbp!]
   \includegraphics[width=1.\linewidth]{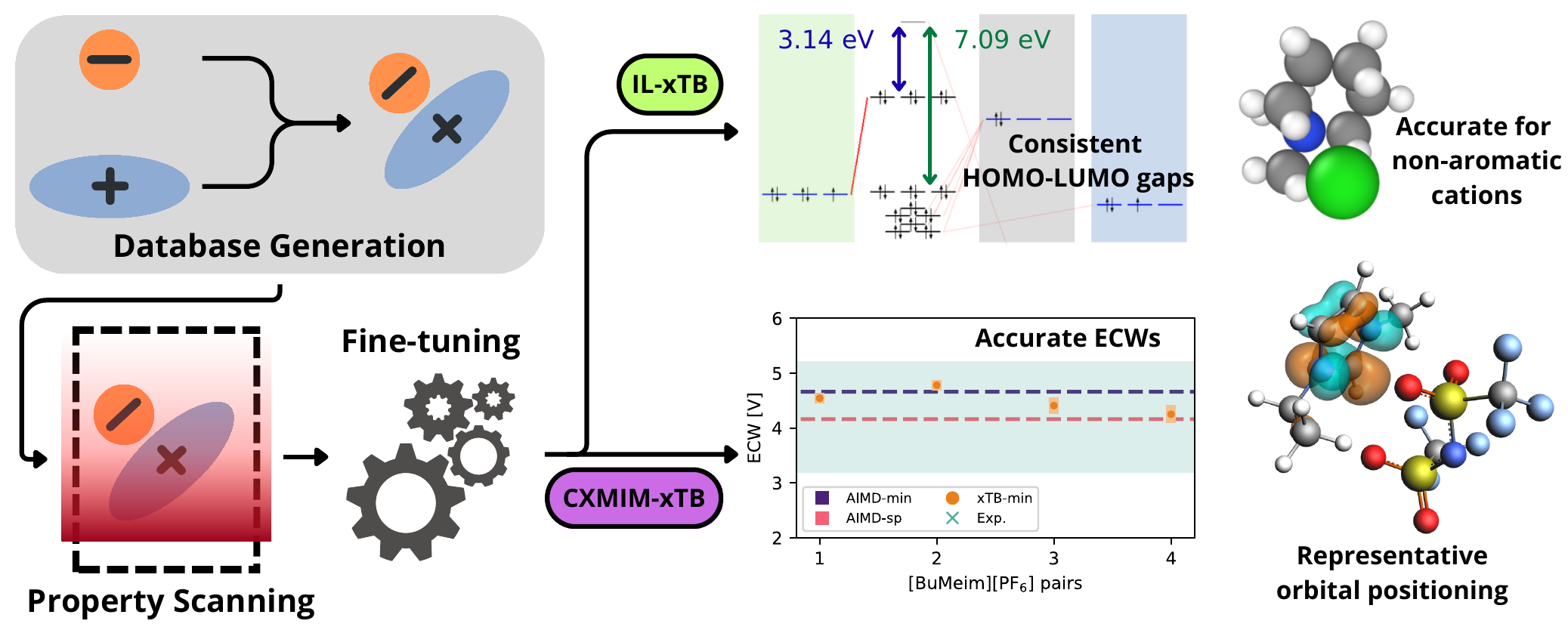}
\end{figure}

%Comment the following lines is SI is not needed

\newpage

\renewcommand{\figurename}{Figure.}
\renewcommand{\thetable}{S\arabic{table}}
\setcounter{table}{0}
\renewcommand{\thefigure}{S\arabic{figure}} 
\setcounter{figure}{0}
\renewcommand{\thesection}{S\arabic{section}}
\setcounter{section}{0}
\renewcommand{\thepage}{S\arabic{page}}
\setcounter{page}{0}

\onecolumngrid

\begin{center}
    \textbf{\Huge{Supplementary Information}}

    \vspace{1.0cm}

    for

    \vspace{1.0cm}

    \textbf{\Large{Reparameterizing GFN1-xTB for Ionic Liquids}}
\end{center}

   \newpage

\section*{Supporting Methods}

%\subsection{SI-M1. Chemical space definition}

%The chemical space considered in this work is summarized in Figure 2 of the main text. It includes nitrogen-based cation families (dialkyl-imidazolium, dialkyl-piperidinium, dialkyl-pyrrolidinium, alkyl-pyridinium) and tetraalkyl-ammonium cations, combined with a set of nine anions.

%For completeness, the full list of cations and anions included in the dataset is provided in Table S1. In total, 55 cations and 9 anions were considered. Not all possible alkyl-chain combinations were included due to the absence of structures in the PubChem database and to limit computational cost. Only one conformation per isolated ion was used.

\subsection{Dataset generation details}

Ion structures were obtained from the PubChem database \cite{10.1093/nar/gkae1059}. The structures were individually optimized at the PBE-D3(BJ) level of theory using the ADF engine of AMS 2024.1 \cite{10.1063/5.0258496}, with DZP basis sets \cite{https://doi.org/10.1002/jcc.10255}. Geometry optimizations were performed using the \texttt{good} convergence criteria. Random ion-pair conformations were generated using the Dokoka code (\url{https://github.com/stflavian/Dokoka}). For each anion-cation combination, 10 configurations were generated by randomly rotating one ion around the geometric center of the other and applying a radial displacement to avoid atomic overlaps within a 2.5~\AA\ radius. The initial displacement vector had a magnitude of 0.1~\AA. Each generated configuration was processed through a three-step geometry optimization protocol:
\begin{enumerate}
    \item UFF pre-optimization using the ForceField engine of AMS \cite{doi:10.1021/ja00051a040},
    \item GFN1-xTB optimization using the standard parameterization \cite{doi:10.1021/acs.jctc.7b00118} with D3(BJ) corrections,
    \item PBE-D3(BJ) optimization using the ADF engine, the large frozen-core approximation, and the FIRE optimizer \cite{PhysRevLett.97.170201}.
\end{enumerate}

Normal convergence criteria were used for all optimization steps except for the initial isolated-ion optimizations. This workflow results in three geometries per initial configuration (UFF, GFN1-xTB, and PBE-D3(BJ)), yielding a total of 30 geometries per ion pair. Representative examples of the structural evolution along this workflow are shown in Figure S1. The full dataset is available at \url{https://doi.org/10.5281/zenodo.20539350} under the CC0 License.

\subsection{Reference calculations and property extraction}

\begin{figure}[t!]
    \centering
    \includegraphics[width=0.5\linewidth]{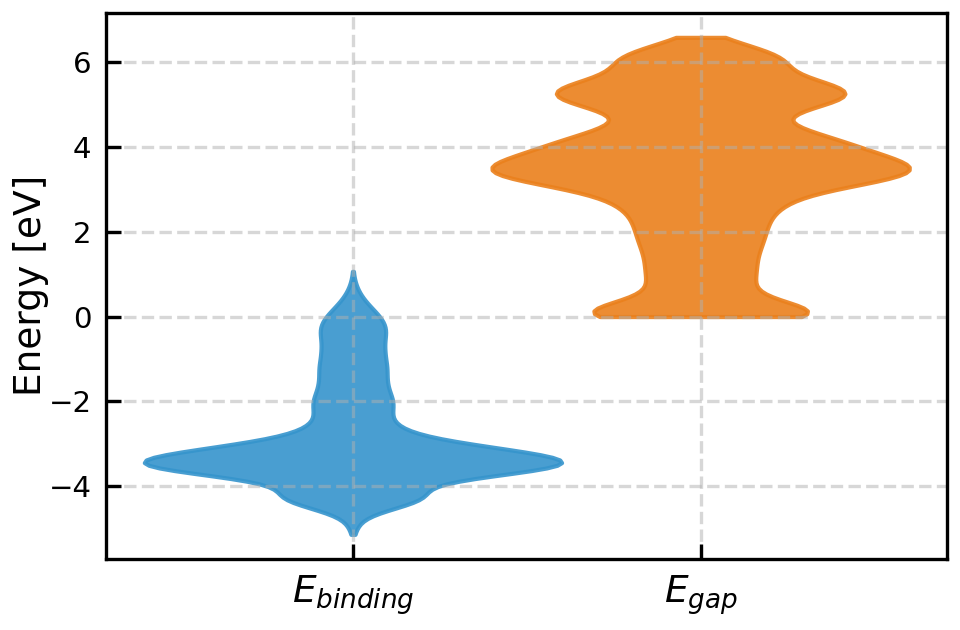}
    \caption{Distribution of binding energies and HOMO-LUMO gaps in the dataset.}
    \label{fig:Fig_S2}
\end{figure}

Single-point calculations were performed for all geometries using the PBE-D3(BJ) level of theory with the ADF engine of AMS. Default settings were used unless otherwise specified. The large frozen-core approximation was employed, and energy gradients were explicitly requested. Binding energies were computed as:

\begin{equation}
E_{\mathrm{bind}} = E_{\mathrm{pair}} - E_{\mathrm{cation}} - E_{\mathrm{anion}}
\end{equation}
where all terms correspond to PBE-D3(BJ) energies. These values are approximate, as basis set superposition error (BSSE) corrections were not included. Atomic forces were obtained from the negative energy gradients. To limit dataset size, forces were only included for selected atomic species depending on the parameterization. The HOMO-LUMO gap was extracted directly from the electronic structure of each system. The distributions of binding energies and HOMO-LUMO gaps in the dataset are shown in Figure \ref{fig:Fig_S2}. Additional properties available in the AMS \texttt{.rkf} files, such as dipole moments and atomic charges, were not included in the training dataset to limit computational cost.

\begin{figure*}[t!]
    \centering
    \includegraphics[width=1\linewidth]{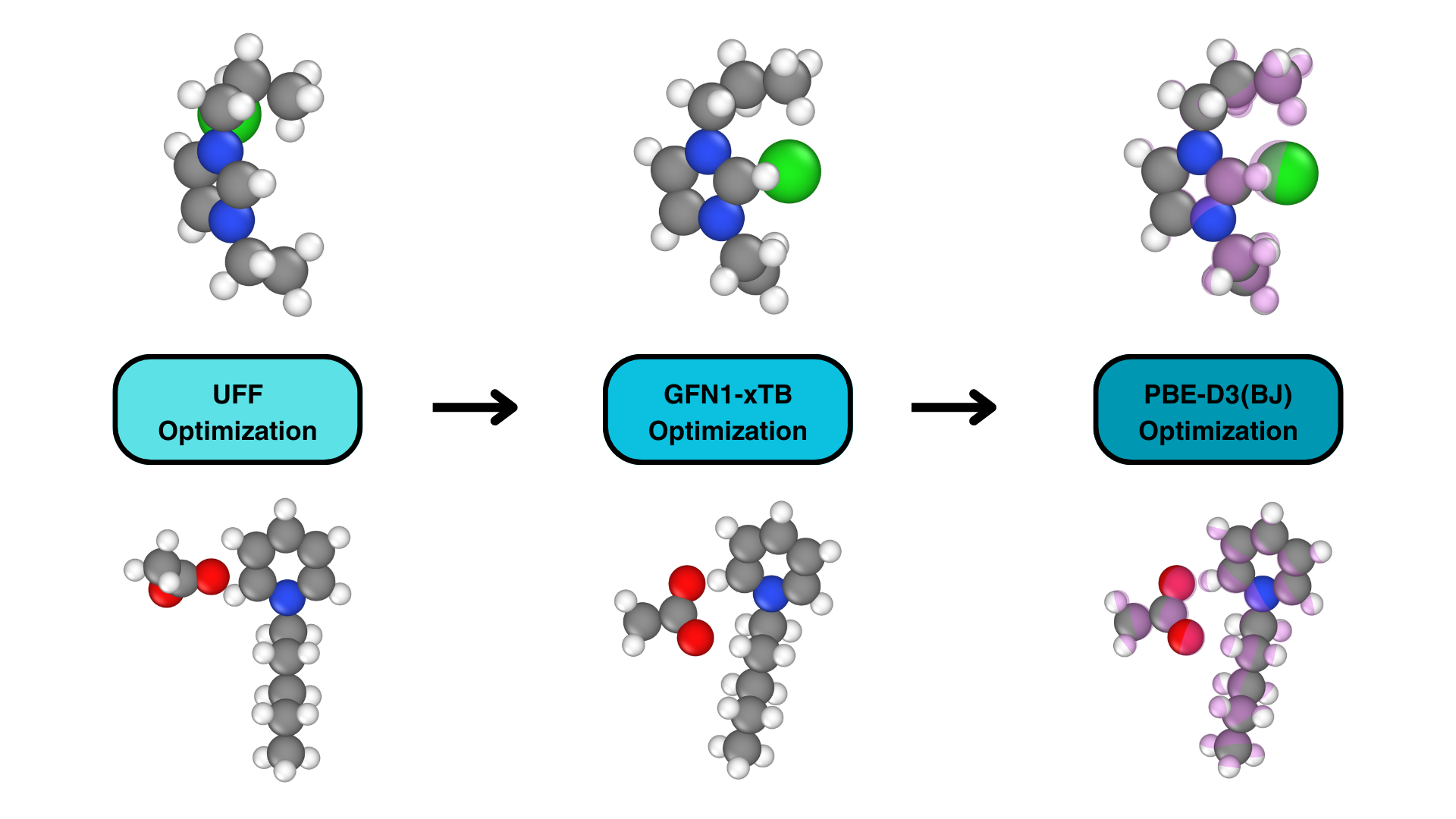}
    \caption{Representative evolution of ionic-pair geometries during the optimization workflow, reproduced from the main dataset generation procedure. The structures undergo successive optimization at the UFF, GFN1-xTB, and PBE-D3(BJ) levels.}
    \label{fig:Fig_S1}
\end{figure*}

\subsection{Parameterization setup}

Parameterization was performed using the ParAMS module \cite{doi:10.1021/acs.jcim.1c00333}. Input files were generated using the \textit{params} Python package. The job collection was defined as a \texttt{.yaml} file containing all systems used in training. Each entry included the system geometry and a unique identifier linking it to the reference data. Geometries were extracted from the PBE-D3(BJ) single-point \texttt{.rkf} files. The training dataset consisted of entries defined by:

\begin{itemize}
    \item system identifier,
    \item property (energy, forces, RMSE, bandgap),
    \item weight $w$,
    \item scaling factor $\sigma$,
    \item unit conversion (if required).
\end{itemize}

Default ParAMS extractors were used. The weight $w$ controls the contribution of each property to the objective function, while $\sigma$ defines the characteristic error scale. The parameter interface was manually defined by activating selected GFN1-xTB parameters and restricting their allowed ranges to improve convergence. Optimization was performed using the CMA-ES algorithm \cite{542381}. The objective function was defined as the root-mean-squared error (RMSE). Training was carried out in batches of 256 jobs, randomly shuffled at each iteration. The population size per generation was set equal to the number of available cores.

The weight value $w$ was determined based on the optimization strategy used for the ion pair, with values of 0.3, 0.6, and 1.0 for UFF, GFN1-xTB, and PBE-D3(BJ) optimized frames respectively. A weight of 1.0 was assigned to individual ions. The scaling factor $\sigma$ was set to 0.1 for all entries.

\newpage

\section*{Supporting images}

\begin{figure*}[h]
    \centering
    \includegraphics[width=0.9\linewidth]{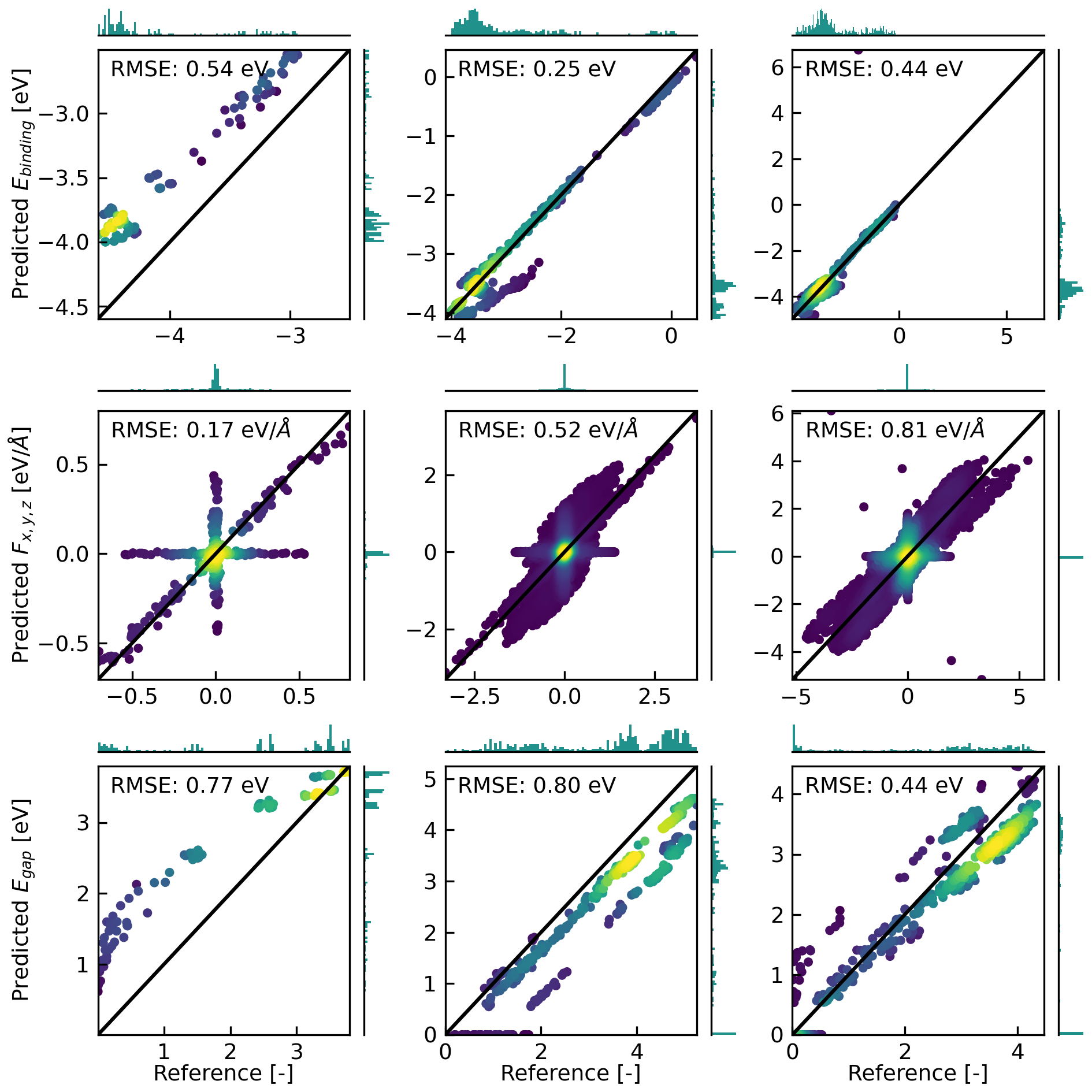}
    \caption{Dataset performance of GFN1-xTB on ionic liquids containing imidazolium-based cations. The three columns represent the results for chloride, fluorinated anions, and oxygen-containing anions from left to right.}
    \label{fig:Fig_S3}
\end{figure*}

\begin{figure*} [t!]
    \centering
    \includegraphics[width=1\linewidth]{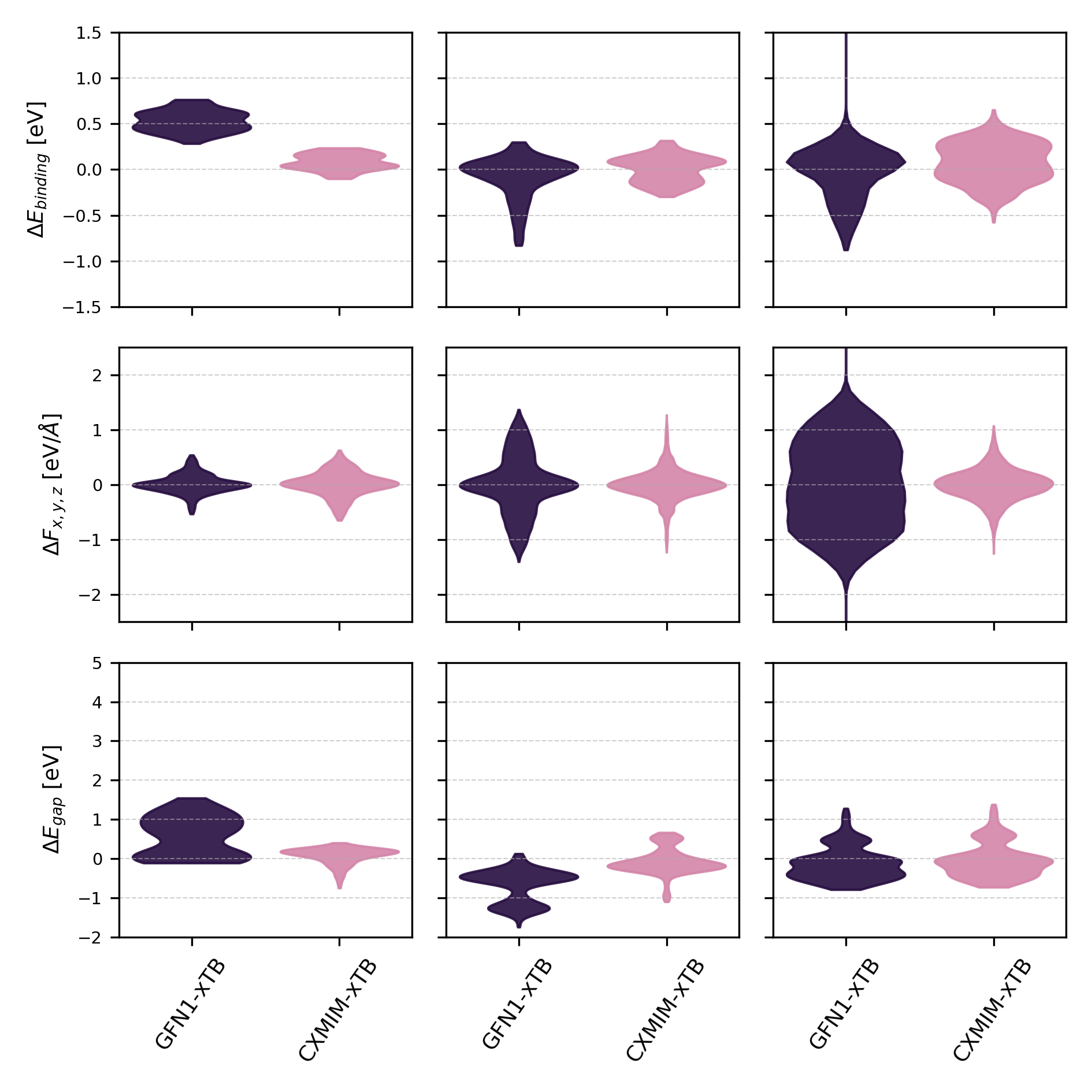}
    \caption{Distribution of errors in GFN1-xTB-computed and CXMIM-xTB-computed binding energies, forces, and HOMO-LUMO gaps compared to PBE-D3(BJ). The three columns show the results for the chloride, fluorinated compounds, and oxygen containing compounds respectively.}
    \label{fig:Fig_S4}
\end{figure*}

\begin{figure*} [t!]
    \centering
    \includegraphics[width=1\linewidth]{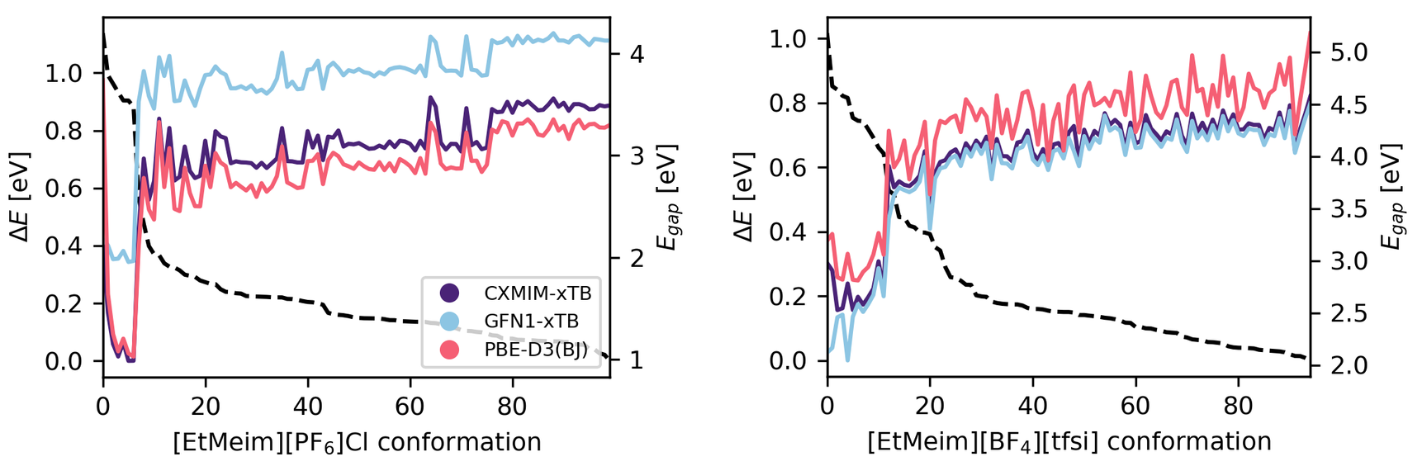}
    \caption{Comparison of HOMO-LUMO gaps calculated for \ch{[EtMeim][PF6]Cl} (left) and \ch{[EtMeim][BF4][tfsi]} (right) clusters using PBE-D3(BJ), GFN1-xTB, and CXMIM-xTB. The left axis measures the relative energy of a cluster with respect to the lowest-energy cluster found. The right axis measures the HOMO-LUMO gap of a cluster.}
    \label{fig:Fig_S5}
\end{figure*}

\begin{figure*}[t!]
    \centering
    \includegraphics[width=0.9\linewidth]{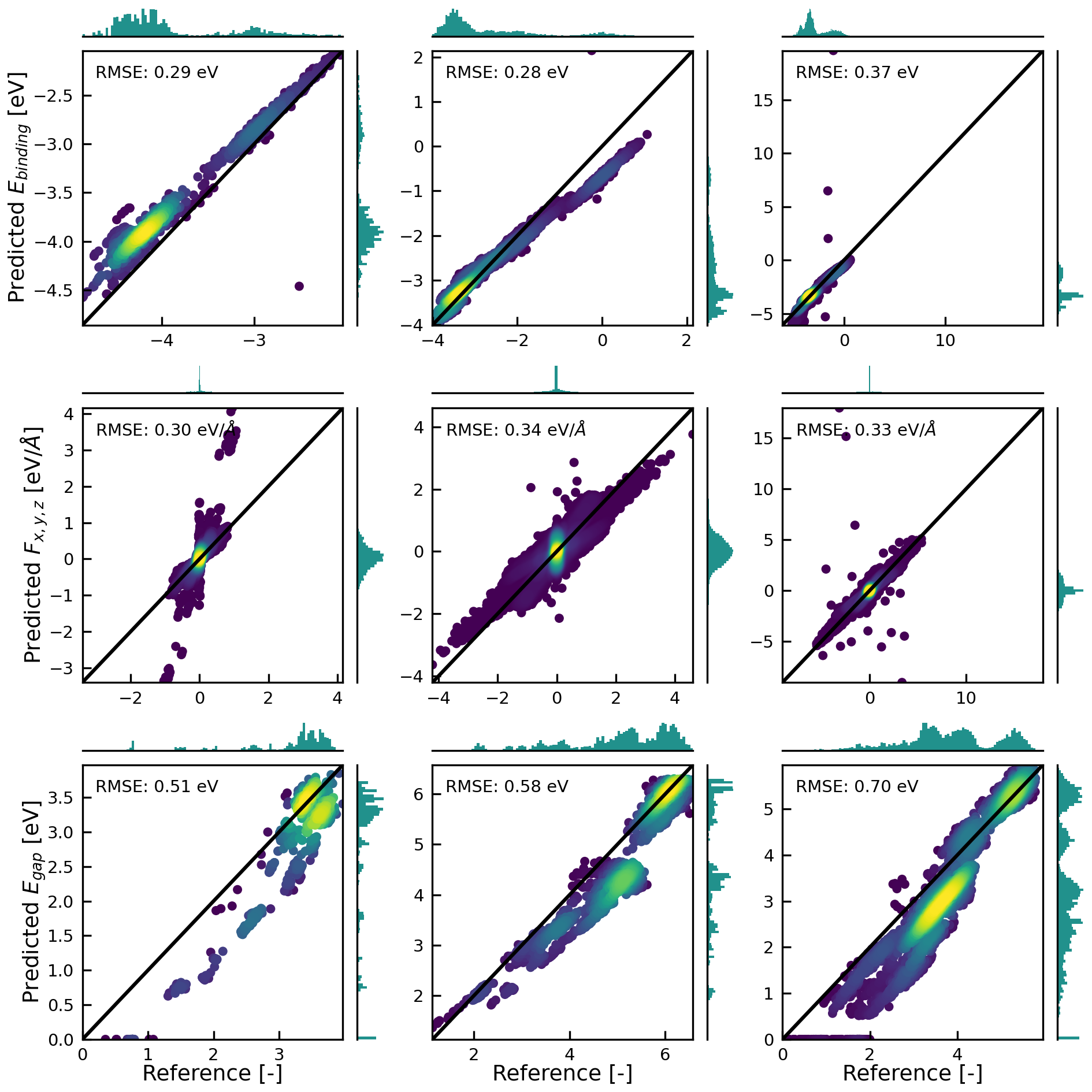}
    \caption{Dataset performance of IL-xTB. The three rows represent the results for chloride, fluorinated anions, and oxygen-containing anions from top to bottom.}
    \label{fig:Fig_S6}
\end{figure*}

\begin{table} [t!]
    \centering
    \caption{CXMIM-xTB parameters.}
    \begin{tabular}{cc}
        Cl &   \\
        \hline
        \hline
        $\eta$ & 0.709791 \\
        $\Gamma$ & 1.007064 \\
        $\alpha$ & 1.332040 \\
        $Z$ & 18.033875 \\
        EN & 3.183429 \\
        & \\
        F &   \\
        \hline
        \hline
        $\eta$ & 0.776963 \\
        $\Gamma$ & 1.661740 \\
        $\alpha$ & 2.363806 \\
        $Z$ & 7.044084 \\
        EN & 3.971171 \\
        & \\
        O &   \\
        \hline
        \hline
        $\eta$ & 0.575441 \\
        $\Gamma$ & 0.010607 \\
        $\alpha$ & 1.935375 \\
        $Z$ & 5.159904 \\
        EN & 3.443300 \\
    \end{tabular}
    \quad
    \begin{tabular}{cc}
        3s &  \\
        \hline
        \hline
        & \\
        $k^{\mathrm{poly}}$ & -9.472479 \\
        $\kappa$ & -1.388976 \\
        $H$ & -22.420630 \\
        $\zeta$ & 2.215361 \\
        & \\
        2s &  \\
        \hline
        \hline
        & \\
        $k^{\mathrm{poly}}$ & -4.942954 \\
        $\kappa$ & -0.869219 \\
        $H$ & -24.343522 \\
        $\zeta$ & 2.990057 \\
        & \\
        2s &  \\
        \hline
        \hline
        & \\
        $k^{\mathrm{poly}}$ & -13.397142 \\
        $\kappa$ & 0.514994 \\
        $H$ & -24.190835 \\
        $\zeta$ & 2.295430 \\
    \end{tabular}
    \quad
    \begin{tabular}{cc}
        3p &  \\
        \hline
        \hline
        & \\
        $k^{\mathrm{poly}}$ & -12.072799 \\
        $\kappa$ & -2.886112 \\
        $H$ & -16.156432 \\
        $\zeta$ & 1.841460 \\
        & \\
        2p &  \\
        \hline
        \hline
        & \\
        $k^{\mathrm{poly}}$ & -11.399009 \\
        $\kappa$ & -1.145201 \\
        $H$ & -18.175429 \\
        $\zeta$ & 2.186495 \\
        & \\
        2p &  \\
        \hline
        \hline
        & \\
        $k^{\mathrm{poly}}$ & -4.650931 \\
        $\kappa$ & 0.163899 \\
        $H$ & -17.608263 \\
        $\zeta$ & 2.083712 \\
    \end{tabular}
    \quad
    \begin{tabular}{cc}
        3d &  \\
        \hline
        \hline
        & \\
        $k^{\mathrm{poly}}$ & 12.708448 \\
        $\kappa$ & 1.193506 \\
        $H$ & -3.436638 \\
        $\zeta$ & 1.660009 \\
        & \\
        3d &  \\
        \hline
        \hline
        & \\
        $k^{\mathrm{poly}}$ & - \\
        $\kappa$ & - \\
        $H$ & - \\
        $\zeta$ & - \\
        & \\
        3d &  \\
        \hline
        \hline
        & \\
        $k^{\mathrm{poly}}$ & - \\
        $\kappa$ & - \\
        $H$ & - \\
        $\zeta$ & - \\
    \end{tabular}
    \label{tab:cxmim_parameters}
\end{table}

\begin{table} [t!]
    \centering
    \caption{IL-xTB parameters.}
    \begin{tabular}{cc}
        Cl &   \\
        \hline
        \hline
        $\eta$ & 0.723884 \\
        $\Gamma$ & 1.0 \\
        $\alpha$ & 0.879760 \\
        $Z$ & 15.748820 \\
        EN & 3.16 \\
        & \\
        F &   \\
        \hline
        \hline
        $\eta$ & 0.697593 \\
        $\Gamma$ & 1.615037 \\
        $\alpha$ & 2.362163 \\
        $Z$ & 6.855032 \\
        EN & 3.98 \\
        & \\
        O &   \\
        \hline
        \hline
        $\eta$ & 0.721878 \\
        $\Gamma$ & -0.005102 \\
        $\alpha$ & 1.893615 \\
        $Z$ & 5.164391 \\
        EN & 3.44 \\
        & \\
        C &   \\
        \hline
        \hline
        $\eta$ & 0.479988 \\
        $\Gamma$ & 1.053856 \\
        $\alpha$ & 1.281954 \\
        $Z$ & 4.428763 \\
        EN & 2.55 \\
    \end{tabular}
    \quad
    \begin{tabular}{cc}
        3s &  \\
        \hline
        \hline
        & \\
        $k^{\mathrm{poly}}$ & -12.746169 \\
        $\kappa$ & -0.877849 \\
        $H$ & -24.824232 \\
        $\zeta$ & 1.808094 \\
        & \\
        2s &  \\
        \hline
        \hline
        & \\
        $k^{\mathrm{poly}}$ & -3.021950 \\
        $\kappa$ & -0.165288 \\
        $H$ & -26.576183 \\
        $\zeta$ & 3.467630 \\
        & \\
        2s &  \\
        \hline
        \hline
        & \\
        $k^{\mathrm{poly}}$ & -13.425889 \\
        $\kappa$ & -0.619379 \\
        $H$ & -22.866927 \\
        $\zeta$ & 2.123836 \\
        & \\
        2s &  \\
        \hline
        \hline
        & \\
        $k^{\mathrm{poly}}$ & -8.459360 \\
        $\kappa$ & 0.076813 \\
        $H$ & -15.397876 \\
        $\zeta$ & 2.039572 \\
    \end{tabular}
    \quad
    \begin{tabular}{cc}
        3p &  \\
        \hline
        \hline
        & \\
        $k^{\mathrm{poly}}$ & -10.798196 \\
        $\kappa$ & -4.043772 \\
        $H$ & -12.450133 \\
        $\zeta$ & 2.036764 \\
        & \\
        2p &  \\
        \hline
        \hline
        & \\
        $k^{\mathrm{poly}}$ & -10.164741 \\
        $\kappa$ & -0.384879 \\
        $H$ & -16.192241 \\
        $\zeta$ & 2.275353 \\
        & \\
        2p &  \\
        \hline
        \hline
        & \\
        $k^{\mathrm{poly}}$ &  -5.476362 \\
        $\kappa$ & 0.351864 \\
        $H$ & -15.814735 \\
        $\zeta$ & 2.230846 \\
        & \\
        2p &  \\
        \hline
        \hline
        & \\
        $k^{\mathrm{poly}}$ & 0.394627 \\
        $\kappa$ & 0.597227 \\
        $H$ & -9.321153 \\
        $\zeta$ & 1.820104 \\
    \end{tabular}
    \quad
    \begin{tabular}{cc}
        3d &  \\
        \hline
        \hline
        & \\
        $k^{\mathrm{poly}}$ & 13.088867 \\
        $\kappa$ & 0.0 \\
        $H$ & -1.190095 \\
        $\zeta$ & 1.932463 \\
        & \\
        3d &  \\
        \hline
        \hline
        & \\
        $k^{\mathrm{poly}}$ & - \\
        $\kappa$ & - \\
        $H$ & - \\
        $\zeta$ & - \\
        & \\
        3d &  \\
        \hline
        \hline
        & \\
        $k^{\mathrm{poly}}$ & - \\
        $\kappa$ & - \\
        $H$ & - \\
        $\zeta$ & - \\
        & \\
        3d &  \\
        \hline
        \hline
        & \\
        $k^{\mathrm{poly}}$ & - \\
        $\kappa$ & - \\
        $H$ & - \\
        $\zeta$ & - \\
    \end{tabular}
    \label{tab:il_parameters}
\end{table}

\begin{table*} [t!]
    \centering
    \caption{Comparison between the accuracy of GFN1-xTB, CXMIM-xTB and IL-xTB with respect to PBE-D3(BJ) reference data.}
    \begin{tabular}{cccccc}
        Atomic & Metric & GFN1-xTB & CXMIM-xTB & GFN1-xTB & IL-xTB \\
        species & RMSE & \ch{[XMeim]+} only & \ch{[XMeim]+} only & complete & complete \\
        \hline
        \hline
        Cl & $E_{binding}$ [eV] & 0.54 & 0.12 & 0.43 & 0.29 \\
        Cl & $F_{x,y,z}$ [eV/\AA] & 0.17 & 0.23 & 0.20 & 0.30 \\
        Cl & $E_{gap}$ [eV] & 0.77 & 0.22 & 2.55 & 0.51 \\
        \hline
        F & $E_{binding}$ [eV] & 0.25 & 0.14 & 0.30 & 0.28 \\
        F & $F_{x,y,z}$ [eV/\AA] & 0.52 & 0.27 & 0.52 & 0.34 \\
        F & $E_{gap}$ [eV] & 0.80 & 0.34 & 0.89 & 0.58 \\
        \hline
        O & $E_{binding}$ [eV] & 0.44 & 0.23 & 0.32 & 0.37 \\
        O & $F_{x,y,z}$ [eV/\AA] & 0.81 & 0.28 & 0.79 & 0.33 \\
        O & $E_{gap}$ [eV] & 0.44 & 0.44 & 1.32 & 0.70 \\
    \end{tabular}
    \label{tab:rmse}
\end{table*}

\clearpage

\section*{Performance benchmark}

To test the performance of the two methods, a workstation equipped with a 32-Core AMD Ryzen Threadripper PRO 5975WX CPU and 128 GB of RAM is used. To avoid issues with memory and thread allocations, no other programs are running in parallel during the calculations. For a complete overview, both single-point and geometry optimization routines are considered; the former uses a \ch{[BuMeim][OAc]} conformation that is pre-optimized using PBE-D3(BJ) to ensure convergence of the calculations, while the latter uses a randomly sampled \ch{[N_{1,1,1,1}]Cl} conformation that was pre-optimized using UFF. The root-mean-squared-displacement (RMSD) between the UFF-optimized conformation and the one of PBE-D3(BJ) and GFN1-xTB is 0.3824 \AA\ and 0.3818 \AA\ respectively. Meanwhile, the RMSD between the optimized conformations of PBE-D3(BJ) and GFN1-xTB is 0.0184 \AA, suggesting that the two resulting geometries are almost identical.

\begin{figure} [h]
    \centering
    \includegraphics[width=0.9\linewidth]{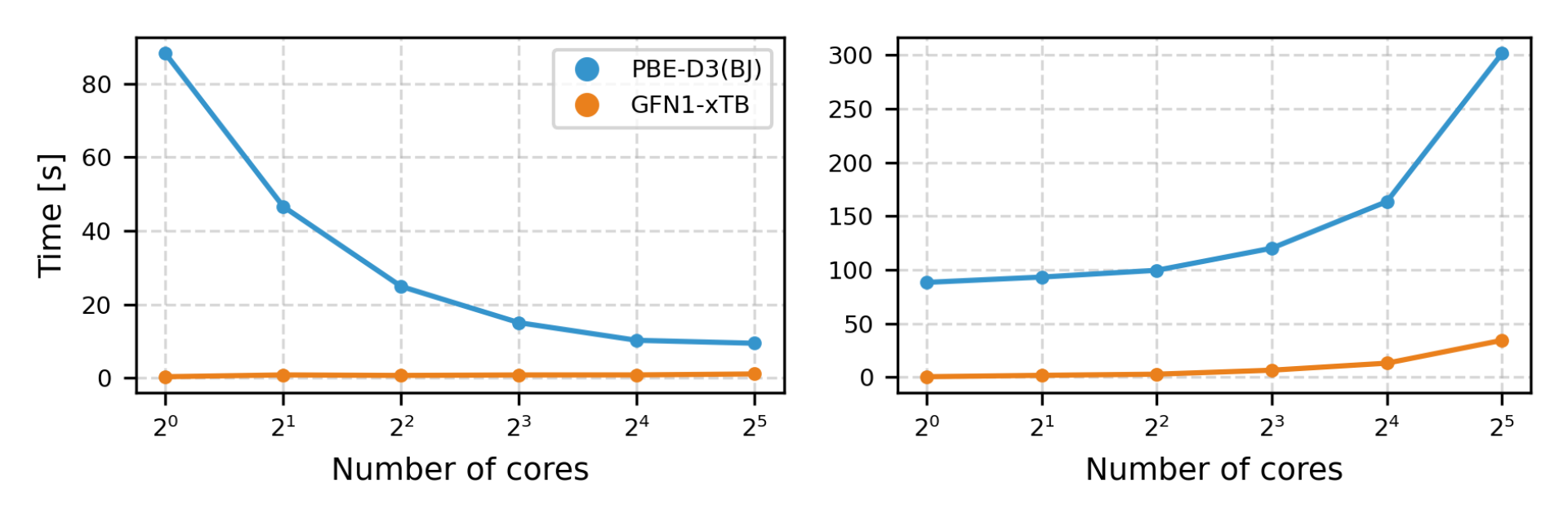}
    \caption{Computational time comparison between PBE-D3(BJ) and GFN1-xTB as a function of the number of CPU cores. The left panel shows the time required for a single-point calculation, highlighting the scalability of PBE-D3(BJ) and the consistently low cost of GFN1-xTB. The right panel displays the cumulative time to complete a batch of 32 single-point calculations under parallel execution. While PBE-D3(BJ) benefits from multi-core acceleration per job, GFN1-xTB is significantly more efficient when leveraging high-throughput parallelism across single-core runs. Moreover, the speedup gained by PBE-D3(BJ) is not sufficient to compensate for the loss in available parallel resources.}
    \label{fig:Fig_S7}
\end{figure}

In the case of single-point calculations (Figure \ref{fig:Fig_S7}), PBE-D3(BJ) shows a reduction in computational time as the number of CPU cores increases. GFN1-xTB, on the other hand, performs best when running on a single core and shows significant slowdown when parallelism is used, going from 0.3 seconds per simulation to 1.1 seconds. As such, ignoring the amount of resources allocated and only comparing the two best performing setups, the semi-empirical method is found to be one order of magnitude faster. 

However, this comparison is misleading when batches of simulations need to be performed, since idle cores on a CPU can be used to run parallel simulations. To account for such a scenario, the results for the single run can be extrapolated; assuming a set of 32 jobs that need to be performed, one can perform either one batch of 32 simulations on one core, two batches of 16 simulations on two cores, and so on. As such, each simulation time can be multiplied by the number of batches required to finish the set of jobs. Performing this analysis (right panel of Figure \ref{fig:Fig_S7}), both PBE-D3(BJ) and GFN1-xTB are found to perform best when maximizing the number of jobs running in parallel, with the speedup gained from running on multiple cores being unable to compensate for the loss in resources. While individual jobs would perform significantly faster, running 32 jobs using PBE-D3(BJ) on 32 cores in series would result in a waiting time approximately three times higher compared to the single-core scenario. Moreover, due to its low resource usage and high speed, GFN1-xTB is found to be almost two orders of magnitude faster compared to its DFT counterpart. These results are found to be consistent with actual GFN1-xTB parallel runs, with the average time of 32 jobs running in parallel on one core being 0.36 seconds, similar to the previous result of 0.32 seconds. 

\begin{figure} [h]
    \centering
    \includegraphics[width=0.9\linewidth]{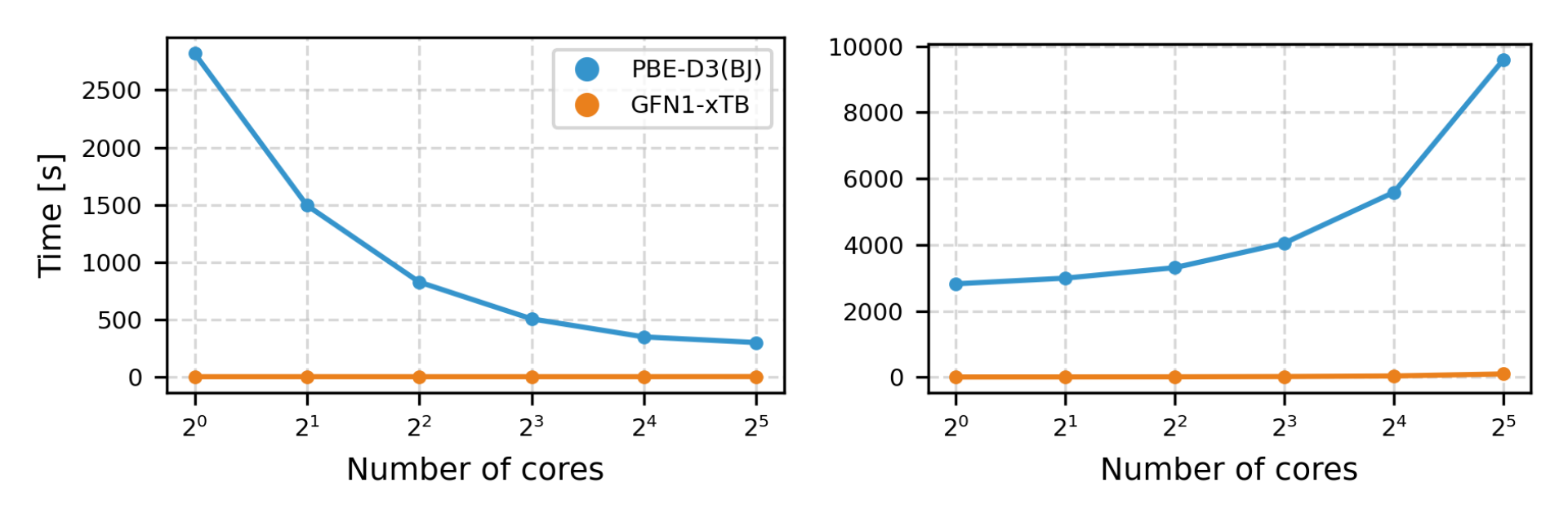}
    \caption{Computational time comparison between PBE-D3(BJ) and GFN1-xTB as a function of CPU cores for geometry optimizations. The left panel shows the time per optimization, following a similar trend to Figure \ref{fig:Fig_S7}. The right panel shows the cumulative time for a batch of 32 optimizations under parallel execution. As before, both methods benefit from running more jobs in parallel; however, GFN1-xTB is approximately three orders of magnitude faster than PBE-D3(BJ) for geometry optimizations.}
    \label{fig:Fig_S8}
\end{figure}

A similar behavior is also observed for geometry optimizations (Figure \ref{fig:Fig_S8}), where the performance of PBE-D3(BJ) is found to improve with increasing number of cores, while GFN1-xTB is found to perform best on a single core. However, the difference in order of magnitude goes up, with the two best performing setups being split by two orders of magnitude (2 seconds for GFN1-xTB against 300 seconds for PBE-D3(BJ)), while the performance on sets of 32 jobs is apart by three orders of magnitude. These results highlight the primary benefit of using DFTB: speed. Given the time required for both methods to run, it is clear that using the semi-empirical model can offer a considerable speed boost to screening studies, allowing for more conformations to be considered in a shorter amount of time.

\end{document}